\documentclass[useAMS,usenatbib, 
]{mnras}
\usepackage{natbib}
\usepackage{graphicx}
\usepackage{epsfig}
\usepackage{amssymb}
\usepackage{amsmath}
\usepackage{natbib}
\usepackage{times}

\usepackage{bm}

\usepackage[english]{babel}

\title[Dust streaming instability in turbulent medium]{The resonant drag instability of dust streaming in turbulent protoplanetary disc}
\author[V.V. Zhuravlev]{V. V. Zhuravlev$^{1}$\thanks{E-mail:
zhuravlev@sai.msu.ru} \\
$^{1}$Sternberg Astronomical Institute, Lomonosov Moscow State University, Universitetskij pr., 13, Moscow 119234, Russia}

\begin{document}

\date{
}

\pagerange{\pageref{firstpage}--\pageref{lastpage}} \pubyear{2017}

\maketitle

\label{firstpage}

\defcitealias{zhuravlev-2019}{Z19}

\begin{abstract}

Damping of the previously discovered resonant drag instability (RDI) of dust streaming in protoplanetary disc 
is studied using the local approach to dynamics of gas-dust perturbations in the limit of the small dust fraction. 
Turbulence in a disc is represented by the effective viscosity and diffusivity in equations of motion for gas and dust, 
respectively. 
In the standard case of the Schmidt number (ratio of the effective viscosity to diffusivity) $Sc=1$, 
the reduced description of RDI in terms of the inertial wave (IW) and 
the streaming dust wave (SDW) falling in resonance with each other reveals that damping solution 
differs from the inviscid solution simply by adding the characteristic damping frequency to its growth rate. 
RDI is fully suppressed at the threshold viscosity, which is estimated analytically, first, for radial drift, next, 
for vertical settling of dust, and at last, in the case of settling combined with radial drift of the dust. 
In the last case, RDI survives up to the highest threshold viscosity, with a greater excess for smaller solids.
Once $Sc\neq 1$, a new instability specific for dissipative perturbations on the dust settling background emerges. 
This instability of the quasi-resonant nature is referred to as settling viscous instability (SVI).
The mode akin to SDW (IW) becomes growing in a region of long waves provided that $Sc>1$ ($Sc<1$). 
SVI leads to an additional increase of the threshold viscosity.

\end{abstract}

\begin{keywords}
hydrodynamics --- accretion, accretion discs --- instabilities --- protoplanetary discs --- planet formation
\end{keywords}

\section{Introduction}

The early evolution of protoplanetary disc is associated with the settling of dust to the disc midplane. 
This process leads to the formation of the high-density thin layer of centimetre- to metre-sized particles. 
Such a layer is subject to the streaming instability of \citet{youdin-goodman-2005} operating due to the substantial 
radial drift of the particles. The streaming instability is a necessary mechanism supporting the growth of gravitationally 
unstable dust clumps on the way to the formation of planetesimals, which has been confirmed in a number of simulations of dust dynamics in laminar discs, 
see \citet{johansen-2007-apj-2}, \citet{bai-stone-2010}, \citet{yang-johansen-2014}, \citet{johansen-2017}.
However, discs are generally turbulent. How the intrinsic turbulence of gas component affects the streaming instability 
was mostly investigated for the particular case of magnetorotational instability-driven turbulence. Simulations 
performed by \citet{fromang-papaloizou-2006}, \citet{johansen-2007-nat}, \citet{balsara-2009} and \citet{tilley-2010}
showed that solids with the dimensionless stopping time $\tau \lesssim 0.1$ do not exhibit strong clumping in turbulent
environment. 
Recently, \citet{umurhan-2019} argued that the initial conditions used in the aforementioned studies are not consistent with either 
the radial-drift or fragmentation barriers of the particle evolution. Particularly, solids with $\tau\gtrsim 0.1$ 
should be destructed by mutual collisions of particles induced by the turbulent gas motion. 
Additionally, they showed that the linear growth rate of the streaming instability in the allowed range of the parameters 
is highly reduced with the account of turbulent viscosity and diffusivity. 
Just recently, \citet{lin-2020} confirm that it is rather difficult to make work the streaming instability in global models of
turbulent protoplanetary discs.

Meanwhile, the local analysis of gas-dust dynamics performed by \citet{squire_2018} revealed that 
the bulk settling of dust considered on the 
length-scale shorter than the disc scale-height and the time-scale shorter than the characteristic time of settling to the disc 
midplane is also linearly unstable.
Moreover, they realised that this new settling instability along with the known streaming instability belongs to a wide class
of instabilities of dust streaming through the gas, which were referred to as resonant drag instabilities (RDI).
The growth rate of the settling instability, as considered in laminar disc,
is independent of $\tau$. Accordingly, it does not reduce for small particles. 
Next, it operates at larger length-scales as compared with the streaming instability:
the characteristic length-scale of growing modes is the streaming velocity of dust times the Keplerian dynamical time, 
whereas in a thin disc, 
the settling velocity can be much higher than the velocity of radial drift of the dust. By comparing the eddy turnover time 
with the growth time of the settling instability, \citet{squire_2018} concluded that settling instability better resists the damping
action of turbulence rather than streaming instability, especially, for lower level of turbulence.

In this situation, it is worth considering the dust clumping due to the settling instability at the stage preceding 
the formation of a thin and dense sub-disc of large particles subject to the streaming instability. 
One of important issues to address on that way is the role of turbulence in the linear growth of the dust density perturbations. 
For that, this study aims to consider turbulence analytically by adding an effective terms describing mean turbulent flux of 
momentum in equation of motion for gas and mean turbulent diffusion of solids in the continuity equation for dust 
in the framework of two-fluid approximation of gas-dust dynamics. 
Modification of the growth rate of the settling instability affected by 
dissipative processes is consistently derived in the leading order in the small dust fraction for the general case of 
dust streaming both vertically and radially. 
The problem is considered also in the leading order in $\tau \ll 1$, 
which corresponds to dynamics in the terminal velocity approximation. For reference, the similar exercise is done 
with the streaming instability, 
which means that dynamics of the radially drifting dust
is considered up to the next order in $\tau$, which is beyond the terminal velocity approximation, 
see \citet{zhuravlev-2019} (\citetalias{zhuravlev-2019} hereafter).

Methodologically, this work is a sequel to \citetalias{zhuravlev-2019} who showed that RDI of the dust settling in laminar
protoplanetary disc arises due to the coupling of the inertial wave (IW) with the streaming dust wave (SDW) 
in the vicinity of the corresponding 
mode crossing, where the waves fall in resonance with each other. The coupling of IW and SDW gives birth to the coupled modes
having the identical pattern speeds and equal growing/damping rates in the finite band of wave-lengths.
The general equations for local dynamics of gas-dust mixture in a disc taken from \citetalias{zhuravlev-2019} 
are supplemented by dissipative terms effectively describing turbulent transfer of mass and momentum. 
Further change to equations for the centre-of-mass velocity of gas-dust mixture and the relative velocity of dust and gas 
allows one to identify the new dimensionless parameter, which defines the order of dissipative terms. 
This is a product of the particle stopping time and the characteristic damping frequency of perturbations, $\omega_\nu$. 
It turns out that as soon as this product is small, what is typical for discs,
the complicated dispersion equation for modes of dissipative gas-dust perturbations
reduces to the form similar to the inviscid variant of the dispersion equation.
Precisely, the coupling term of the dispersion equation saves its inviscid form, while dissipative corrections are contained
in the separate dispersion equations for IW and SDW. In the standard case when there is no difference in magnitudes of
the effective viscosity and diffusivity, the damping rates of IW and SDW in the absence of the dust backreaction on gas are identical to $\omega_\nu$, 
whereas the mode coupling retains its inviscid form except that the growth (damping) rate of the coupled mode
representing the band of RDI is decreased (increased) by $\omega_\nu$. Employing this simple solution, 
it can be shown analytically that RDI persists up to the highest level of turbulence when the dust streams {\it almost} vertically
with a small portion of radial drift. It is estimated that in this case RDI vanishes at the threshold viscosity 
by $\tau^{-3/4}$ times higher than the threshold viscosity of the streaming instability of \citet{youdin-goodman-2005}.
Clearly, this is a significant advantage for small particles especially bearing in mind that
the growth rate of the settling instability tends to its inviscid value at smaller viscosity, whereas its 
inviscid growth rate is by $\tau^{-1}$ times higher 
than that of the streaming instability, see \citet{squire_2018} and \citetalias{zhuravlev-2019}.

The second part of this paper is focused on the more general situation of viscosity differing from diffusivity.
The particular case of the diminished diffusivity is considered in more detail, so that the Schmidt number, 
which is defined here as the ratio of the effective viscosity to diffusivity of gas
\footnote{In this paper the turbulent diffusivity of solids is assumed to be equal to that of the gas, since $\tau$ is small, 
see \citet{youdin-lithwick-2007}.}, $Sc>1$. The numerical simulations of turbulence provide the evidence that 
this may be the case in protoplanetary discs, see e.g. \citet{carballido-2005} who obtained $Sc\sim 11$
studying the dust radial diffusion, or \citet{johansen-klahr-mee-2006} whose $Sc$ associated with vertical diffusion of dust 
was larger than 10. Also, \citet{fromang-papaloizou-2006} found $Sc=2.8$ in their study of the dust settling 
in the presence of turbulence.

It is revealed that as $Sc > 1$, 
SDW and IW producing the settling instability decouple from each other, 
which means that they acquire different pattern speeds inside the band of the 
corresponding inviscid RDI. At the same time, the decrement of the less damping (more damping) mode 
outside of the band of the corresponding inviscid RDI decreases (increases)
{\it excessively} to what is expected for $Sc>1$ merely from the difference between the damping rates of SDW and IW in the absence 
of the dust backreaction on gas. 
Thus, as $Sc>1$, there is an additional destabilisation of the turbulent gas-dust mixture comparing to the case $Sc=1$. 
In the limit $Sc\to \infty$, the less damping mode, which turns out to be akin to SDW, 
becomes growing for any value of
viscosity in the {\it unbounded} range of wave-lengths. 
It is shown that there is an optimal value of viscosity, which 
causes the maximum destabilisation of the system with respect to the case $Sc=1$. 
Since the inviscid SDW is a negative energy wave, which was shown by \citetalias{zhuravlev-2019}, this additional instability 
emerging for $Sc>1$ is similar to the viscous instability, see e.g. \citet{cairns-1979} and \citet{nezlin-1976}. 
Accordingly, it is referred to as settling viscous instability (SVI) below.
For finite $1<Sc<\infty$ SVI, which operates around the band of the inviscid RDI, 
still vanishes at some threshold viscosity, though, higher than that for $Sc=1$. 
It may be important that for viscosity under this threshold value SVI extends far into the region of long wavelengths. 
Hence, there are particular modes akin to SDW with wavelengths much larger than the typical length-scale of RDI, which 
are growing only in the turbulent medium with $Sc > 1$. In contrast to situation near the resonance between IW and SDW, 
those modes become stable once $Sc\to 1$.

At last, it is checked that as $Sc<1$, the overall picture of SVI remains the same. However, the growth is provided by 
the mode akin to IW rather than SDW. An explanation of this feature is given in Summary.

\section{Local dynamics of gas-dust mixture with effective viscosity of gas and diffusivity of dust}

\subsection{General equations}

In this work the model used by \citetalias{zhuravlev-2019} is generalised in order to take into account 
the effective viscosity of gas and the effective diffusivity of dust.
The variables and notations are adopted from \citetalias{zhuravlev-2019}.
The notations for frequently used variables are summarised in the Appendix \ref{app_symb}.  
The Cartesian coordinates $x,y,z$ stand, respectively, for radial $r$, azimuthal $\varphi$ and
vertical directions in the vicinity of some point $r_0,\varphi_0, z_0>0$ above the disc midplane. 
The reference frame is rotating with angular velocity $\Omega_0$ around the host star. It is assumed that 
$\Omega_0$ is equal to the angular velocity of disc at $r=r_0$.
In order to consider the local subsonic perturbations it is sufficient to take a small domain
$\{x,y,z\}\ll h\ll r_0$, where $h$ is the disc scaleheight. 
Dynamics of two components of turbulent gas-dust mixture is described by Euler equations for gas

\begin{equation}
\label{eq_U_g}
\begin{aligned}
{\cal D} {\bf U}_g - 2\Omega_0 U_{g,y} {\bf e}_x + (2-q)\Omega_0 U_{g,x} {\bf e}_y + \\
({\bf U}_g \nabla){\bf U}_g = - \frac{\nabla p}{\rho_g} + \nu \nabla^2 {\bf U}_g + \frac{\rho_p}{\rho_g} \frac{\bf V}{t_s},
\end{aligned}
\end{equation}
and dust 
\begin{equation}
\label{eq_U_p}
\begin{aligned}
{\cal D} {\bf U}_p - 2\Omega_0 U_{p,y} {\bf e}_x + (2-q)\Omega_0 U_{p,x} {\bf e}_y + \\
({\bf U}_p\nabla){\bf U}_p = \frac{\nabla p_0}{\rho_g} - \frac{{\bf V}}{t_s},
\end{aligned}
\end{equation}
see also \citet{umurhan-2019}.
Eqs. (\ref{eq_U_g}-\ref{eq_U_p}) are coupled to each other through aerodynamic drag, 
which is the last term in the right-hand side (RHS) of both equations. The aerodynamic drag is parametrised
by the particles' stopping time $t_s=const$. 
Equations (\ref{eq_U_g}-\ref{eq_U_p}) are formulated for gas and dust velocities, ${\bf U}_g$ and ${\bf U}_p$, respectively. 
A relative velocity of dust with respect to gas environment is
${\bf V}\equiv {\bf U}_p -{\bf U}_g$. The time derivative ${\cal D} \equiv \partial_t - q\Omega_0 x \partial_y$
contains an advective term since 
it is assumed that ${\bf U}_{g,p}$ are measured with respect to reference single-fluid rotating shear flow introduced 
by the toroidal velocity ${\bf U}_0 \equiv -q\Omega_0 x {\bf e}_y$ and the corresponding pressure, $p_0$ and density, $\rho_g$. 
The non-zero viscosity does not change ${\bf U}_0$ on the local scale.
Similar to the inviscid case, the addition to gas pressure, $p$, arises due to the interaction of gas with dust. 
The dust mass density is denoted by $\rho_p$.
Note that throughout this work it is assumed that mixture is not dominated by dust, so that the dust fraction 
\begin{equation}
\label{f}
f\equiv \frac{\rho_p}{\rho_g} < 1.
\end{equation}
The effective damping force, which arises due to intrinsic turbulent motions of gas, is proportional to kinematic viscosity coefficient, $\nu$, assumed to be constant hereafter. 
In the local approach the Laplacian $\nabla^2 \equiv \partial_{xx} + \partial_{yy} + \partial_{zz}$.
In the case of subsonic perturbations eq. (\ref{eq_U_g}) is accompanied by the condition of the divergence-free motion of gas
\begin{equation}
\label{eq_rho}
\nabla \cdot {\bf U}_g = 0.
\end{equation}

As has been discussed in the literature, the turbulent diffusion of dust can be appropriately described by 
the following kinetic term 
added to RHS of the continuity equation for dust,
\begin{equation}
\label{term_diff}
D \nabla \cdot \left [ \rho_g \nabla \left ( \frac{\rho_p}{\rho_g} \right ) \right ],
\end{equation}
see e.g. \citet{morfill-1984}, \citet{dubrulle-1995}, \citet{fromang-papaloizou-2006} and \citet{charnoz-2011}.
As far as in the local model considered here $\rho_g$ is assumed to be constant in both stationary and perturbed solutions, 
what is expressed by eq. (\ref{eq_rho}),
the continuity equation for dust reads
\begin{equation}
\label{eq_sigma}
{\cal D} \rho_p + \nabla (\rho_p{\bf U}_p) = D \nabla^2 \rho_p.
\end{equation}
The intensity of dust diffusion is defined by the diffusion coefficient $D$.

Further, eqs. (\ref{eq_U_g}), (\ref{eq_U_p}), (\ref{eq_rho}) and (\ref{eq_sigma}) 
yield the following equation for the center-of-mass velocity,
$$
{\bm U} \equiv \frac{\rho_g{\bm U}_g + \rho_p {\bm U}_p}{\rho},
$$
\begin{equation}
\label{eq_U}
\begin{aligned}
{\cal D} {\bf U} - 2\Omega_0 U_y {\bf e}_x + (2-q) \Omega_0 U_x {\bf e}_y + ({\bf U}\nabla) {\bf U} + \\ 
\frac{\rho_g}{\rho} \left \{  \left ( {\bf V} \nabla \left ( \frac{\rho_p}{\rho} \right ) \right ) {\bf V} +
2\frac{\rho_p}{\rho} \left ( {\bf V} \nabla \right ) {\bf V} \right \} 
= \\
\frac{\nabla p_0}{\rho_g} - \frac{\nabla(p+p_0)}{\rho} + \frac{\rho_g}{\rho} \left ( \nu \nabla^2 {\bf U}_g + 
D \frac{{\bf V}}{\rho} \nabla^2 \rho_p \right ),
\end{aligned}
\end{equation}
where the last term in the RHS accounts for both effective viscosity of gas and effective diffusion of dust and 
$\rho\equiv \rho_g+\rho_p$ is the total density of gas-dust mixture. 
Equation for the relative velocity is the following
\begin{equation}
\label{eq_V}
\begin{aligned}
{\cal D} {\bf V} 
- 2\Omega_0 V_y {\bf e}_x + (2-q) \Omega_0 V_x {\bf e}_y + \\ ({\bf U}\nabla) {\bf V}  + ({\bf V}\nabla) {\bf U} + 
\frac{\rho_g}{\rho} ({\bf V} \nabla) \left ( \frac{\rho_g}{\rho} {\bf V} \right ) - \\ 
\frac{\rho_p}{\rho} ({\bf V} \nabla) \left ( \frac{\rho_p}{\rho} {\bf V} \right ) 
 =  \frac{\nabla (p+p_0)}{\rho_g} - \frac{\rho}{\rho_g} \frac{{\bf V}}{t_s} - \nu \nabla^2 {\bf U}_g.
\end{aligned}
\end{equation}

Eqs. (\ref{eq_U}-\ref{eq_V}) accompanied by the continuity equation
\begin{equation}
\label{eq_rho_tot}
{\cal D} \rho_p + \nabla ( \rho {\bf U} ) = D \nabla^2 \rho_p,
\end{equation}
along with the condition (\ref{eq_rho}), where 
\begin{equation}
\label{U_g_V}
{\bf U}_g = {\bf U} - \frac{\rho_p}{\rho} {\bf V},
\end{equation}
describe the local dynamics of turbulent gas-dust mixture in a disc.
They are identical to eqs. (11-14) of \citetalias{zhuravlev-2019} in the case of laminar disc.

\subsection{Terminal velocity approximation for gas-dust dynamics with dissipative effects}

There are general restrictions, which make it possible to deal with the reduced description 
of gas-dust mixture local dynamics.
The reduced description neglects with the inertia of solids, 
what is referred to as the terminal velocity approximation (TVA hereafter), 
see \citet{youdin-goodman-2005}.
The corresponding equations follow from eqs. (\ref{eq_U}-\ref{eq_V}), where the terms non-linear in ${\bf V}$ and 
the rest of the terms in the left-hand side (LHS) of eq. (\ref{eq_V}) are omitted, explicitly 
\begin{equation}
\label{eq_U_2}
\begin{aligned}
{\cal D} {\bf U} - 2\Omega_0 U_y {\bf e}_x + (2-q) \Omega_0 U_x {\bf e}_y + ({\bf U}\nabla) {\bf U} = \\ 
\frac{\nabla p_0}{\rho_g} - \frac{\nabla(p+p_0)}{\rho} + \frac{\rho_g}{\rho} \left ( \nu \nabla^2 {\bf U}_g  + 
D \frac{{\bf V}}{\rho} \nabla^2 \rho_p \right ),
\end{aligned}
\end{equation}

\begin{equation}
\label{eq_TVA}
\frac{\nabla (p+p_0)}{\rho} = \frac{{\bf V}}{t_s} + \nu \frac{\rho_g}{\rho} \nabla^2 {\bf U}_g
\end{equation}
supplemented by eqs. (\ref{eq_rho}), (\ref{eq_rho_tot}) and (\ref{U_g_V}).


Let there be small parameters of the problem
\begin{equation}
\label{tva_1}
\tau_* \equiv t_s \max\{ t_{ev}^{-1},\Omega_0 \} \ll 1
\end{equation}
and 
\begin{equation}
\label{tva_2}
\lambda^{-1} \equiv \frac{l_s}{l_{ev}} \ll 1,
\end{equation}
where $t_{ev}$ and $l_{ev}$ are, respectively, the characteristic time- and length-scales of the problem, while
\begin{equation}
\label{l_s}
l_s \equiv g t_s^2
\end{equation}
is the stopping length of solids defined by the absolute value of the specific pressure gradient\footnote{In the reference frame
rotating with $\Omega_0$ gas and dust feel the star gravity in both vertical and radial directions in a disc 
with a caveat that in the latter case gravity is effective as being mostly balanced by centrifugal force.}
$$
g \equiv \left| \frac{\nabla (p + p_0)}{\rho} \right |.
$$
For the growing perturbations considered in this study, it is sufficient to use 
the restriction (\ref{tva_1}) in a more particular form, 
\begin{equation}
\label{tva_1_add}
\tau \equiv t_s \Omega_0 \ll 1 
\end{equation}
as the growth timescales are larger than the dynamical timescale.

Then, various terms omitted in eqs. (\ref{eq_U_2}-\ref{eq_TVA}) are as small as $\sim \tau$, 
or $\sim \sqrt{\lambda^{-1}}$, or $\sim \lambda^{-1}$ 
with respect to $g$, see Section 2.1 of \citetalias{zhuravlev-2019}. 
As it is discussed there, in the case when the gradient term in LHS of eq. (\ref{eq_U}) $|({\bf U \nabla}){\bf U}|\sim U^2$,
it dominates the rest of terms there being 
the zeroth order term\footnote{In certain cases, when the inertial terms 
in LHS of eq. (\ref{eq_U}) become dominant, one finds that $U \sim \Omega_0^{-1} g$, while the gradient terms 
$\sim UV$ in LHS of eq. (\ref{eq_V}) become of the order of $\lambda^{-1}/\tau$, which does not change the main conclusions about
the conditions of TVA.} 
in both $\tau$ and $\lambda^{-1}$ along with the second term in RHS of eq. (\ref{eq_V}) $\sim V$.
Therefore, the following order-of-magnitude relations are generally valid
$$
V \sim g t_s, \quad U \sim \frac{g t_s}{\sqrt{\lambda^{-1}}},
$$
so that the dissipative terms $\sim U$ and $\sim V$ in RHS 
of eqs. (\ref{eq_U_2}-\ref{eq_TVA}) are, respectively, of the order of $\tau_{\nu}/\sqrt{\lambda^{-1}}$ and 
$\tau_{\nu}$ as compared with $g$. Here 
\begin{equation}
\label{tau_nu}
\tau_{\nu} \equiv t_s \frac{\nu}{l_{ev}^2}
\end{equation}
is the dimensionless stopping time in units of the characteristic time of damping of gas motion through the action of viscous forces.

In a turbulent disc, $\tau_{\nu}$ can be of the order of unity or even much larger than unity.
Indeed, the specific pressure gradient can be as high as vertical gravitational acceleration estimated as
$
g \sim \Omega_0^2 h.
$
For the given $t_s$ this puts the most stringent low limit on $l_{ev}$ following from eq. (\ref{tva_2}). 
The marginal restriction on $l_{ev} \gtrsim l_s$ yields
\begin{equation}
\label{tau_nu_est}
\tau_{\nu} \lesssim \frac{\alpha}{\tau^3},
\end{equation}
where it is assumed that coefficient of gas effective viscosity is parametrised according to \citet{shakura-sunyaev-1973}
\begin{equation}
\label{visc_alpha}
\nu = \alpha \Omega_0 h^2.
\end{equation}
It is generally accepted that $\alpha\ll 1$ in protoplanetary discs, however,
$\tau_{\nu}$ can take any value as compared with unity due to its strong inverse dependence on $\tau$, see eq. (\ref{tau_nu_est}).
For this reason, the full dissipative terms are retained in eqs. (\ref{eq_U_2}-\ref{eq_TVA}).
It should be noted that length-scale corresponding to $\tau_\nu \sim 1$ is quite a small value in a disc
\begin{equation}
\label{l_ev_nu}
l_{ev} (\tau_\nu\sim 1) \sim (\tau \alpha)^{1/2} h.
\end{equation}
However, it can be much larger than the stopping length, since
$$
\frac{l_{ev} (\tau_\nu\sim 1) } {l_s (z_0\sim h)} \sim \frac{\alpha^{1/2}}{\tau^{3/2}}. 
$$

At the same time, the dust diffusivity is considerable within TVA as far as $l_D \gtrsim l_s$, where 
$l_D$ is defined as the length-scale, where the RHS of eq. (\ref{eq_rho_tot}) is comparable to the first term 
in LHS therein. For the prescription (\ref{visc_alpha}) this implies that
\begin{equation}
\label{D_est}
(\alpha/Sc)^{1/2} \gtrsim \tau^2,
\end{equation}
where $Sc \equiv \nu / D$ is the Schmidt number generally not equal to unity, while it is assumed that $\Omega_0 t_{ev} \sim 1$.
For $Sc \sim 1$ eq. (\ref{D_est}) gives a weaker restriction on the low value of $\alpha$ than the one following from 
eq. (\ref{tau_nu_est}) for $\tau_\nu \sim 1$. However, it is clear that dissipative effects should strongly influence 
the dynamics of gas-dust mixture for small particles with $\tau \ll 1$.

\subsection{Stationary streaming of dust}

The bulk settling and radial drift of the dust is described locally by the following stationary solution of 
eqs. (\ref{eq_U}), (\ref{eq_V}), (\ref{eq_rho}) and (\ref{eq_rho_tot})
\begin{equation}
\label{bg_U}
{\bf U} = 0,
\end{equation}
\begin{equation}
\label{bg_p}
\frac{\nabla (p+p_0)}{\rho} = - g_x {\bf e}_x - g_z {\bf e}_z,
\end{equation}
\begin{equation}
\label{bg_V}
{\bf V} = -t_s ( g_x {\bf e}_x + g_z {\bf e}_z) + \frac{\kappa^2}{2\Omega_0} \frac{t_s^2}{1+f} g_x {\bf e}_y,
\end{equation}
\begin{equation}
\label{bg_sigma}
\rho_p=const
\end{equation}
where $\kappa^2 \equiv 2(2-q)\Omega_0^2$ is the epicyclic frequency squared, while 
\begin{equation}
\label{g_z}
g_z \equiv -\frac{\partial_z p_0}{\rho_g} \approx \Omega_0^2 z_0
\end{equation}
and 
\begin{equation}
\label{g_x}
g_x \equiv -\frac{\partial_x p_0}{\rho_g} \approx 2 \eta \Omega_0^2 r_0
\end{equation}
with $\eta$ introducing the relative deviation of the disc angular frequency from the Keplerian value.
Both $g_x$ and $g_z$ are assumed to be constant in a small patch of disc. 
The solution (\ref{bg_U}-\ref{bg_sigma}) recovers that of \citetalias{zhuravlev-2019}, 
where the terms of the order of $\sim t_s^3$ and higher have been omitted\footnote{Note that there is a misprint in $V_y$ from 
equation (89) of \citetalias{zhuravlev-2019}, which is fixed here.}. This is a local variant of known \citet{nakagawa-1986} solution.
Therefore, it is not modified by the non-zero viscosity of gas and diffusivity of dust.



Formally, the solution (\ref{bg_U}-\ref{bg_sigma}) exists anywhere in a turbulent disc. However, the study of the dust settling 
in a turbulent medium on scale comparable to the disc scale-height shows that settling is prevented by 
turbulent diffusion because of the global vertical gradient of the dust density. This occurs as far as 
\begin{equation}
\label{restr_stat_1}
\frac{z_0}{h} \lesssim \left ( \frac{\alpha}{\tau} \right )^{1/2},
\end{equation}
see \citet{fromang-papaloizou-2006}. Eq. (\ref{restr_stat_1}) is derived assuming that $Sc\sim 1$. 
For the given $\alpha$ and $z_0$ eq. (\ref{restr_stat_1}) provides the range of $\tau > \tau_{stl}$, where 
$\tau_{stl} \equiv \alpha (z_0/h)^{-2}$, for solids 
that sediment in the presence of turbulent fluctuations. 
Note that the restriction (\ref{restr_stat_1}) can be weakened if dust diffusion is suppressed. This may be the case
when, first, $Sc>1$ and, second, turbulence is inhibited by the dust backreaction on gas.
The latter may occur if turbulence has another origin and properties rather than magnetorotational instability-driven turbulence
simulated by \citet{fromang-papaloizou-2006}.
For example, \citet{lin-2019} showed that the dust backreaction on gas partially suppresses the hydrodynamic turbulence
generated by the vertical shear instability. This is caused by the additional stabilising buoyancy produced by solids, 
see \citet{lin-youdin-2017}.

From the other side, it is possible to estimate $\tau$, when the effective viscosity or/and diffusivity suppress  
RDI caused by the dust settling. The action of the effective viscosity becomes substantial as soon as 
the dissipative term $\sim U$ in eq. (\ref{eq_U}) is of the order of $g$, which yields
\begin{equation}
\label{restr_stat_2}
\alpha \sim \left ( \frac{l_{ev}}{h} \right )^{3/2},
\end{equation}
where it was assumed that $g\sim \Omega_0^2 h$ and $\nu$ is given by eq. (\ref{visc_alpha}).
The action of the effective diffusivity becomes substantial as soon as the dissipative term in RHS of eq. (\ref{eq_rho_tot}) 
is of the order of the first term in LHS of eq. (\ref{eq_rho_tot}), which yields
\begin{equation}
\label{restr_stat_3}
\alpha \sim Sc \left ( \frac{l_{ev}}{h} \right )^2.
\end{equation}
Eq. (\ref{restr_stat_3}) implies that the characteristic damping frequency of perturbations is of order of 
$\Omega_0$, see below.

As it is known from the inviscid analysis, see \citet{squire_2018} and \citetalias{zhuravlev-2019}, 
RDI caused by the dust settling takes place in the vicinity of resonance between SDW and IW given by the condition 
\begin{equation}
\label{l_ev_RDI}
l_{ev} \sim \tau z_0,
\end{equation}
which results in the restrictions that RDI is not damped by turbulence when
\begin{equation}
\label{restr_stat_4}
\tau \gtrsim \alpha^{2/3} \left (\frac{z_0}{h} \right )^{-1}
\end{equation}
and
\begin{equation}
\label{restr_stat_5}
\tau \gtrsim \tau_{dmp} \equiv \left ( \frac{\alpha}{Sc} \right )^{1/2} \left (\frac{z_0}{h} \right )^{-1},
\end{equation}
following, respectively, from eqs. (\ref{restr_stat_2}) and (\ref{restr_stat_3}).
It can be seen that the latter restriction on $\tau$ is stronger than the former one provided that $Sc\sim 1$. 
By this reason, it can be used to 
formulate the overall condition on the physical relevance of the model used in this work. That is, a substantial damping of 
the growing modes of RDI by the effective viscosity and diffusivity should occur prior to termination of the bulk settling of dust 
as one goes from large particles to smaller particles, i.e. $\tau_{stl} < \tau_{dmp}$, or equivalently,
\begin{equation}
\label{restr_stat_6}
\alpha \lesssim \left ( \frac{z_0}{h} \right )^2 Sc^{-1}.
\end{equation}
Once inequality (\ref{restr_stat_6}) is true, both $\tau_{stl}$ and $\tau_{dmp}$ are less than unity, which
is consistent with the basic assumption of smallness of $\tau$ used in this study considering the dynamics of 
particles with $\tau \gtrsim \tau_{dmp}$. 

\subsection{The resonance between axisymmetric modes of dissipative gas-dust perturbations}

Equations (\ref{eq_U}-\ref{eq_rho_tot}) suggest that small
perturbations of gas-dust mixture are described by 
the Eulerian perturbations of the centre-of-mass velocity, ${\bf u}$, the relative velocity, ${\bf v}$, 
the gas pressure, $p^\prime$, and the density of dust, $\rho_p^\prime$.
In this Section, the dispersion equation for axisymmetric modes of perturbations with the account for dissipative
effects is considered in the two different cases.
In the first case, dust is allowed to stream both vertically and radially, whereas the dynamics of perturbations 
is considered within TVA. In the other case, the dust is subject to the radial drift only, however, the dynamics of perturbations
is considered including the next order terms over $\tau$, which are responsible specifically for quasi-resonant modification
of modes akin to SDW and IW leading to classical streaming instability, see the reasoning in \citetalias{zhuravlev-2019}. 
Further, the solutions to these variants of the dispersion equation are compared with each other, see Section \ref{sec_Sc_1}.

\subsubsection{Dispersion equation within TVA}
\label{sec_within_TVA}

Equations (\ref{eq_U_2}), (\ref{eq_TVA}) supplemented by equations (\ref{eq_rho}), (\ref{eq_rho_tot}) and (\ref{U_g_V}) 
have to be linearised on the uniform background given by equations (\ref{bg_U}-\ref{bg_sigma}). 
The corresponding equations for perturbations in the limit of $\tau_\nu \ll 1$ and $f\ll 1$ are derived in the Appendix \ref{app_pert}, see eqs. (\ref{u_x}-\ref{div_u_2}).

It is suitable to write those equations in terms of new variables appropriate for vortical perturbations of velocity:
$\varpi,\, \phi, \, u_z$ and $\delta$,
where
\begin{equation}
\label{chi}
\varpi \equiv -\partial_z u_y, \,\,
\phi \equiv \partial_z u_x
\,\,
\mbox{and}\,\,
\delta \equiv \frac{\rho_p^\prime}{\rho_p}.
\end{equation}
Taking the curl and the divergence of eqs. (\ref{u_x}-\ref{u_z}) one arrives at the following set of equations
\begin{equation}
\label{Sys_1}
\partial_t \phi = \partial_{tx} u_z  - 2\Omega_0 \varpi + f (g_z \partial_x - g_x \partial_z) \delta + \nu \bar\nabla^2 \phi - \nu \bar \nabla^2 \partial_x u_z,  
\end{equation}
\begin{equation}
\label{Sys_2}
\partial_t \varpi = \frac{\kappa^2}{2\Omega_0} \phi + \nu \bar\nabla^2 \varpi,
\end{equation}
\begin{equation}
\label{Sys_3}
\partial_{tx}\varpi = - \frac{\kappa^2}{2\Omega_0} \partial_{zz} u_z + \nu \bar\nabla^2 \partial_x \varpi,
\end{equation}
\begin{equation}
\label{Sys_4}
\partial_{tz}\delta = \tau ( g_z \partial_{zz} + g_x \partial_{xz} ) \delta + 2\Omega_0 \tau \partial_x \varpi + 
D \bar \nabla^2 \partial_z \delta,
\end{equation}
where the term $\propto f\tau$ has been neglected in eq. (\ref{Sys_4}). Eqs. (\ref{Sys_1}-\ref{Sys_4}) recover eqs. (26-29) of 
\citetalias{zhuravlev-2019} in the inviscid limit $\nu=D=0$.

A plane wave of gas-dust perturbations 
\begin{equation}
\label{fourier}
\chi = \hat \chi \exp(-{\rm i}\omega t + {\rm i} {\bf k x}),
\end{equation}
where $\chi \equiv \{\,\varpi,\, \phi, \, u_z,\, \delta\,\}$,
is a particular solution of eqs. (\ref{Sys_1}-\ref{Sys_4}).
Here ${\bf k x } = k_x x + k_z z$, where $k_x$ and $k_z$ are the real wavenumbers, respectively, along local radial and vertical 
directions in a disc, whereas the condition $\Im[\omega]>0$ corresponds to exponentially growing plane waves, 
i.e. to instability of dust streaming through the turbulent gas environment.

There are the following equations for the complex Fourier amplitude, $\hat \chi$:
\begin{equation}
\label{mode_eq_1}
-{\rm i} \omega \hat \phi - \omega k_x \hat u_z + 2\Omega_0 \hat \varpi - {\rm i} f (g_z k_x - g_x k_z) \hat \delta = 
-\nu k^2 \hat \phi + \nu k^2 \, {\rm i}k_x \hat u_z,
\end{equation}
\begin{equation}
\label{mode_eq_2}
-{\rm i} \omega \hat \varpi - \frac{\kappa^2}{2\Omega_0} \hat \phi = -\nu k^2 \hat\varpi,
\end{equation}
\begin{equation}
\label{mode_eq_3}
\omega k_x \hat \varpi - \frac{\kappa^2}{2\Omega_0} k_z^2 \hat u_z = - \nu k^2 \, {\rm i} k_x \hat \varpi,
\end{equation}
\begin{equation}
\label{mode_eq_4}
\omega k_z \hat \delta + \tau g_z k_z^2 \hat \delta + \tau g_x k_x k_z + D k^2{\rm i} k_z \hat \delta - 
2\Omega_0 \tau {\rm i} k_x \hat \varpi = 0,
\end{equation}
where $k^2 \equiv k_x^2 + k_z^2$.

Eqs. (\ref{mode_eq_1}-\ref{mode_eq_4}), yield the following dispersion equation for $\omega$ 
\begin{equation}
\label{disp}
D_g(\omega,{\bm k}) \cdot D_p(\omega,{\bm k}) = \epsilon({\bm k}),
\end{equation}
where
\begin{equation}
\label{D_g}
D_g(\omega,{\bm k}) \equiv (\omega+{\rm i}\omega_\nu)^2 - \omega_i^2,
\end{equation}
\begin{equation}
\label{D_p}
D_p(\omega,{\bm k}) \equiv  \omega + {\rm i} \omega_D - \omega_p, 
\end{equation}
\begin{equation}
\label{eps}
\epsilon({\bm k}) \equiv f t_s \kappa^2 \frac{k_x}{k} \frac{k_z}{k} (k_x g_z - k_z g_x) 
\end{equation}
with $\omega_i\equiv (k_z/k)\kappa$, $\omega_p \equiv - t_s (k_x g_x + k_z g_z)$ and
\begin{equation}
\label{om_nu}
\omega_\nu \equiv \nu k^2,
\end{equation}
\begin{equation}
\label{om_D}
\omega_D \equiv D k^2.
\end{equation}
Equations (\ref{om_nu}-\ref{om_D}) introduce the characteristic frequencies, which describe the rate of, respectively, 
viscous and diffusive damping of modes. In a laminar disc $\omega_\nu \to 0$, $\omega_D \to 0$ and eq. (\ref{disp}) recovers 
the dispersion equation obtained in \citetalias{zhuravlev-2019} within TVA.
In a turbulent disc both $\omega_\nu$ and $\omega_D$ are non-zero. Nevertheless, the dispersion equation for axisymmetric gas-dust 
modes saves its basic form introduced by eq. (\ref{disp}), where LHS is a product of two independent dispersion relations
describing different waves, while RHS, which is proportional to $f t_s$, can be thought of as 
the coupling term, which leads to the mode coupling. The mode coupling occurs in the
vicinity of resonance between the waves, also referred to as the mode crossing, when both $D_g$ and $D_p$ vanish 
at the same frequency and wavevector. 
The concept of the mode coupling was first applied to 
the problem of single fluid dynamics by \citet{cairns-1979}. It was suggested recently for interpretation of RDI of the 
laminar gas-dust mixture by \citetalias{zhuravlev-2019}. 

It is important that the coupling term $\epsilon({\bm k})$ 
is identical to its inviscid counterpart. 
Hence, the dissipative effects 
suppress RDI through the damping of the modes themselves, rather than through modification of coupling between the modes. 
As far as the coupling term is negligible, which is true when $f \to 0$, 
eq. (\ref{disp}) splits into two independent dispersion equations, one of which describes two branches
of damping IW propagating in the opposite directions, 
\begin{equation}
\label{disp_IW}
\omega = \pm \omega_i - {\rm i} \omega_\nu,
\end{equation}
while the other describes the single damping SDW,
\begin{equation}
\label{disp_SDW}
\omega = \omega_p - {\rm i} \omega_D.
\end{equation}
Note that IW is damping solely due to the action of the effective viscosity. On the contrary, 
SDW is damping solely due to the action of the effective diffusivity of dust.



\subsubsection{The reduced dispersion equation beyond TVA}
\label{sec_beyond_TVA}

Here the purpose is to derive the dispersion equation, which describes RDI in the case of the dust streaming only radially,
taking into account the dissipative effects. For that, the derivation 
starts from the full equations (\ref{eq_rho}, \ref{eq_U}-\ref{eq_rho_tot}) linearised on the background given 
by eqs. (\ref{bg_U}-\ref{bg_sigma}) for $g_z=0$. It proceeds along the lines of Section 4.2 of \citetalias{zhuravlev-2019}.
This essentially means that the terms of the higher order in $\tau$ have to be retained in the coupling term, whereas LHS
of the dispersion equation represented by the product of $D_g(\omega,{\bm k})$ and $D_p(\omega,{\bm k})$ remains the same 
as in the previous Section.
The linearised dissipative terms, ${\bf F}_\nu$ and ${\bf G}_\nu$, entering equations 
for, respectively, ${\bf u}$ and ${\bf v}$ are given by eqs. (\ref{F_nu}-\ref{G_nu}). 
Eqs. (\ref{F_nu}-\ref{G_nu}) are valid for dynamics of 
perturbations beyond TVA. Therefore, they are used to add the corresponding new terms to equations describing RDI 
of the dust streaming radially in laminar disc.
As opposed to derivation of the coupling term within TVA, see eq. (\ref{eps}) and the Appendix \ref{app_pert}, 
it is necessary to retain the terms $\sim \tau_\nu$ in eqs. (\ref{F_nu}-\ref{G_nu}).

It can be checked that the only such term 
that leads to modification of the coupling term of the reduced dispersion equation is
\begin{equation}
\label{visc_add_beyond_TVA}
f k^2 (\nu-D) \hat \delta {\bm V}, 
\end{equation}
which has to be added to RHS of the corresponding inviscid equation for ${\bf u}$.  
The term (\ref{visc_add_beyond_TVA}) is constructed of terms $\sim \rho_p^\prime$ 
from the square brackets in eq. (\ref{F_nu}) for ${\bf F}_\nu$.
There are other dissipative contributions to equation for ${\bf u}$ coming from ${\bf F}_\nu$. 
However, they are proportional to $f^2 \hat \delta$, which is not relevant to RDI.
On the other hand, the equation governing the behaviour of the dust density 
does not contain any of dissipative terms $\sim {\bm u}$, since they could appear from the divergence of ${\bf v}$ only.
In turn, the divergence of ${\bf v}$ is derived from equation for ${\bf v}$ which contains ${\bf G}_\nu$.
However, $\nabla \cdot {\bf G}_\nu = 0$ for vortical motion of gas, see equation (\ref{G_nu}).
Changing to the new variables introduced in Section \ref{sec_within_TVA}
one obtains the following dispersion equation in the case of $g_z=0$ (and $f\ll 1$):

\begin{equation}
\label{disp_next_order}
D_g(\omega,{\bm k}) D_p(\omega,{\bm k}) = \tilde \epsilon_\nu(\omega,{\bm k}),
\end{equation}
where the functions in LHS are given by eq. (\ref{D_g}) and (\ref{D_p}), while 
\begin{equation}
\label{t_eps}
\begin{aligned}
\tilde \epsilon_\nu = -t_s f \kappa^2 k_x g_x \frac{k_z^2}{k^2} (1+{\rm i} t_s \omega - 2{\rm i} t_s \omega_p) \\
\left [ 1 + \frac{{\rm i} k_x}{g_x k_z^2} (\omega + {\rm i}\omega_\nu ) (\omega + {\rm i}\omega_D + \omega_p t_s(\omega_\nu-\omega_D) \right .\\ \left . \biggl. 
- {\rm i} t_s \omega_p  + t_s(\omega_\nu-\omega_D) (1 + {\rm i}t_s (\omega +{\rm i} \omega_\nu) ) \right ]
\end{aligned}
\end{equation}
with
\begin{equation}
\label{om_p_red}
\omega_p = - t_s k_x g_x
\end{equation}
specifically for this dispersion equation.

\section{Damping of RDI: crucial role of the dust settling}
\label{sec_Sc_1}

It is known that the growth rate of the inviscid RDI $ \propto f^{1/2}\Omega_0$, see \citet{squire_2018}, 
therefore, for $f\ll 1$ it is considered as a small quantity 
as compared with the rotational frequency.
In order to address the problem of damping of RDI it is sufficient to estimate a degree of dissipation 
when growth rate vanishes, i.e. the damping correction to the inviscid growth rate quenches the growth of perturbations.
The dispersion relations for IW and SDW with the account of dissipation suggest that their damping rates in the absence 
of the mode coupling are given, respectively, by $\omega_\nu$ and $\omega_D$. 
RDI decays as soon as the inviscid value of its growth rate becomes comparable to individual damping rates of 
IW and SDW coupling with each other. Thus, it suffices to consider the effective viscosity and diffusivity, which provide 
$\omega_\nu, \omega_D \sim f^{1/2}\Omega_0$. Since in the vicinity of the mode crossing 
$\|\omega\| \sim \|\omega_p\| \sim \Omega_0$ are of the zeroth order in $f$,
the viscous terms entering $\tilde \epsilon_\nu$ are small compared to the next-order terms in $\tau$ responsible for 
the emergence of RDI. For the purpose of this work those terms will be omitted hereafter. 
Hence, a replacement
\begin{equation}
\label{replace}
\omega \equiv \omega^\prime - {\rm i} \omega_D
\end{equation}
brings eq. (\ref{disp}) to equation
\begin{equation}
\label{disp_mod}
D^\prime_g(\omega^\prime,{\bm k}) \cdot D^\prime_p(\omega^\prime,{\bm k}) = \epsilon({\bm k}),
\end{equation}
while eq. (\ref{disp_next_order}) to another equation
\begin{equation}
\label{disp_next_order_mod}
D^\prime_g(\omega^\prime,{\bm k}) \cdot D^\prime_p(\omega^\prime,{\bm k}) = \tilde \epsilon(\omega^\prime,{\bm k})
\end{equation}
with
$$
\tilde \epsilon \equiv \tilde \epsilon_\nu\, \biggl|_{ \omega_\nu = \omega_D = 0 }.
$$ 

In eqs. (\ref{disp_mod}) and (\ref{disp_next_order_mod})
\begin{equation}
\label{D_g_mod}
D^\prime_g(\omega^\prime,{\bm k}) \equiv (\omega^\prime+{\rm i}\omega_*)^2 - \omega_i^2
\end{equation}
and
\begin{equation}
\label{D_p_mod}
D^\prime_p(\omega^\prime,{\bm k}) \equiv  \omega^\prime - \omega_p, 
\end{equation}
where
\begin{equation}
\label{om_diff}
\omega_* \equiv \nu_* k^2 \equiv \omega_\nu - \omega_D = \omega_\nu (1 - Sc^{-1}).
\end{equation}

In the particular case of $Sc=1$ and $\omega_*=0$ eqs. (\ref{disp_mod}) and (\ref{disp_next_order_mod}) 
become identical 
to their inviscid variants, see, respectively, eq. (49) and eq. (107) of \citetalias{zhuravlev-2019}.
In this case, for both of 
the dispersion equations, the growth rate of RDI with the account of turbulent dissipation is the following 
\begin{equation}
\label{visc_invisc}
\Im[\omega]  = -{\rm i} \omega_\nu + \Delta,
\end{equation}
where $\Delta$ is the corresponding inviscid value of the growth rate determined for either of the two models introduced 
in Sections \ref{sec_within_TVA} and \ref{sec_beyond_TVA}.
The rest of this Section is dedicated to the analysis of the case $Sc=1$, while another plausible situation of $Sc \neq 1$ is considered in Section \ref{sec_Sc_infty}.

The solution (\ref{visc_invisc}) makes it possible to carry out simple estimations of the upper limit of the effective viscosity,
introduced by $\alpha=\alpha_{max}$ according to eq. (\ref{visc_alpha}), when RDI is still operating, $\Im[\omega]>0$. 
The maximum values of the inviscid RDI growth rate of the dust streaming vertically and/or radially were analytically estimated by \citet{squire_2018}. These results were recovered by 
\citetalias{zhuravlev-2019} as the small corrections to the frequencies of both IW and SDW at the crossing of these modes. 
They are employed further in order to obtain $\alpha_{max}$. In what follows, the case of $k_{x,z} > 0$ is considered only. 
According to
the analysis of \citetalias{zhuravlev-2019}, this corresponds to the type-I mode crossing, which is not the only one as dust 
streams both vertically and radially. However, it can be checked that in contrast to the type-I mode crossing, 
the other mode crossings taking place at $k_x<0$ do not have the long length-scale asymptotic, which suggests that they are much more susceptible to dissipative effects. 
The mode crossing is provided by the condition $\omega_p = \omega_i$, 
which can be brought into the form usefull below:
\begin{equation}
\label{res_cond}
\frac{k_x}{\tilde k_x} + \frac{k_z}{\tilde k_z} = \frac{k_z}{k},
\end{equation}
where
\begin{equation}
\label{t_k_x}
\tilde k_{x,z} \equiv \frac{\kappa}{t_s g_{x,z}}.
\end{equation}
In a Keplerian disc, $\tilde k_x$ and $\tilde k_z$ can be estimated using the definitions (\ref{g_z}) and (\ref{g_x}). 
Explicitly,
\begin{equation}
\label{t_k_x_h}
\tilde k_x h \sim \frac{\delta_*}{\eta \tau},
\end{equation}
where the disc aspect ratio is $\delta_* \equiv h/r_0$. As far as the small particles in geometrically thin disc are considered,
$\tilde k_x$ introduces the length-scale much smaller than the disc scale-height.
Further,
\begin{equation}
\label{t_k_z_h}
\tilde k_z z_0 \sim \frac{1}{\tau}.
\end{equation}
As far as $z_0$ is of the order of $h$, and assuming that $\eta \sim \delta_*^2$, eq. (\ref{t_k_z_h}) introduces 
the length-scale greater than $\sim \tilde k_x^{-1}$ by factor of $\delta_*^{-1}$, but still much smaller than the disc scale-height.

\begin{figure}
\begin{center}
\includegraphics[width=8cm,angle=0]{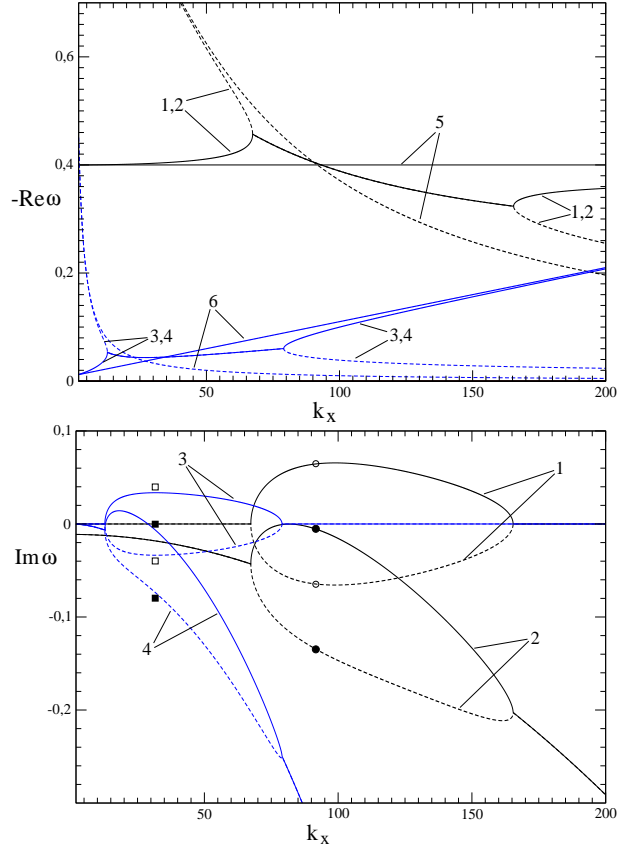}
\end{center}
\caption{
The solution of the dispersion equation (\ref{disp}) for $\tau=0.01$, $q=3/2$ and $Sc=1$.
Top panel: the pattern speed of the modes.
For the curves marked as '5' and '6', the solid and the dashed lines represent, respectively, SDW and IW for $f=0$.
For all other curves the solid and the dashed lines represent the modes akin to, respectively, SDW and IW for $f=0.01$. 
The bottom panel: the growth/damping rate of the modes.
The dashed lines represent the damping coupled mode.
On both panels: the curves marked as '1' and '2' show the mode coupling for $\nu=0$ and $\nu=7\cdot 10^{-6}$, 
respectively, while the other parameters are $g_z=1.0$, $g_x=0$ and $k_z=40$;
the curves marked as '3' and '4' show the mode coupling for $\nu=0$ and $\nu = 4\cdot 10^{-5}$, respectively, while
the other parameters are $g_z=1.0$, $g_x=0.1$ and $k_z=1.0$.
The hollow and the filled circles provide the corresponding analytical estimate, see eq. (\ref{Im_om_vert}), 
of the growth/damping rate at the mode crossing of the solid and the dashed lines marked as '5' obtained for 
$g_z=1.0$, $g_x=0$ and $k_z=40$.
The hollow and the filled squares provide the corresponding analytical estimate, see eq. (\ref{Im_om_rad_vert}), 
of the growth/damping rate at the mode crossing of the solid and the dashed '6' obtained for $g_z=1.0$, $g_x=0.1$ and $k_z=1.0$.
On the coloured versions of plots black and blue curves represent the solutions for, respectively, $k_z=40$ and $k_z=1.0$.
} \label{fig_1}
\end{figure}

\begin{figure}
\begin{center}
\includegraphics[width=8cm,angle=0]{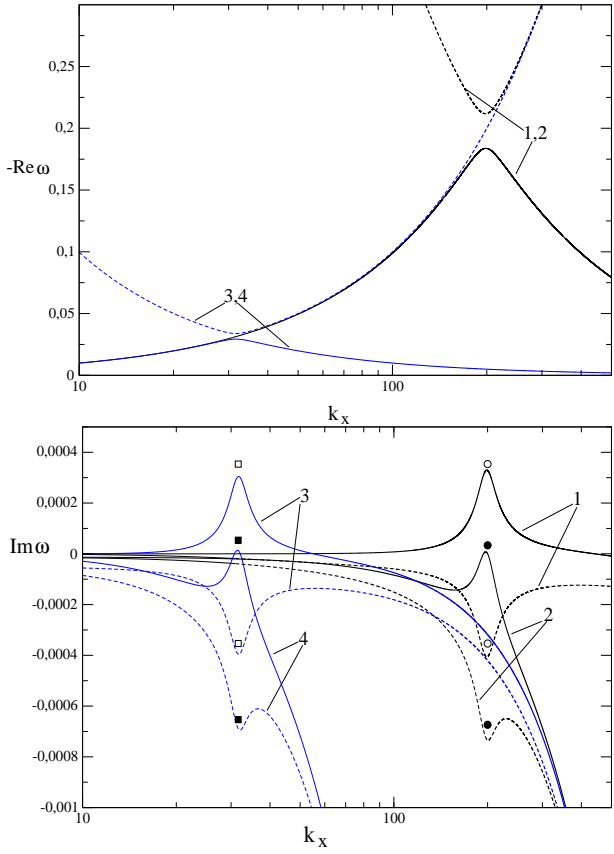}
\end{center}
\caption{
The solution of the dispersion equation (\ref{disp_next_order}) for 
$\tau=0.01$, $f=0.01$, $q=3/2$, $g_z=0.0$, $g_x=0.1$ and $Sc=1$.
Top and bottom panels represent, respectively, the pattern speed and the growth/damping rate of the modes.
The dashed lines represent the solutions, which are damping in the inviscid limit. 
The curves marked as '1' and '2' show the solution for $\nu=0$ and $\nu=8\cdot 10^{-9}$, respectively, while $k_z=40$.
The curves marked as '3' and '4' show the solution for $\nu=0$ and $\nu=3\cdot 10^{-7}$, respectively, while $k_z=1$.
On the coloured versions of plots black and blue curves represent the solutions for, respectively, $k_z=40$ and $k_z=1.0$.
The bottom panel: the hollow and the filled circles provide the corresponding analytical estimate, see eq. (\ref{Im_om_rad}),
of the growth/damping rate at the mode crossing of SDW and IW for $k_z=40$;  
The hollow and the filled squares provide the same analytical estimate  
of the growth/damping rate at the mode crossing of SDW and IW for $k_z=1$.
} \label{fig_2}
\end{figure}

\subsection{Radial drift}
\label{sec_rad_dr}

Close to the disc midplane $g_z\to 0$, $\tilde k_z \to \infty$, and the second term in LHS of eq. (\ref{res_cond}) becomes
small compared to the first one therein. 
Then, according to eq. (\ref{res_cond}) the long wavelength IW and SDW with $k_x \ll \tilde k_x$, 
which are less affected by viscosity/diffusivity, fall in resonance at 
\begin{equation}
\label{res_cond_rad_appr}
k_x \approx (k_z \tilde k_x)^{1/2},
\end{equation}
where $k_z \ll k_x \ll \tilde k_x$. 

There is the following estimation for the maximum value of RDI inviscid growth rate adopted from \citetalias{zhuravlev-2019}
\begin{equation}
\label{gen_approx_sol}
\Delta \approx {\rm i}\, \Omega_0 \tau \left ( \frac{f}{8} \right )^{1/2} \frac{k_z^2}{k^2} 
 \left ( 2 + \frac{k_x^2}{k_z^2} \right ).
\end{equation}

Eq. (\ref{gen_approx_sol}) with eq. (\ref{visc_invisc}) yield in the limit given by eq. (\ref{res_cond_rad_appr}) 
\begin{equation}
\label{Im_om_rad}
\frac{\Im[\omega]}{\Omega_0} \approx \tau \left ( \frac{f}{8} \right )^{1/2} - \alpha h^2 \tilde k_x k_z
\end{equation}
and 
\begin{equation}
\label{alp_max_rad}
\alpha_{max} \approx \left ( \frac{f}{2} \right )^{1/2} \, \frac{z_0}{h}\, \tau^2 \frac{\eta}{(k_z z_0) \delta_*}. 
\end{equation}

Note that it is required that $k_z \gtrsim z_0^{-1}$ here and below in this study in order to stay within the local 
model of disc with the dust settled up to $z \sim z_0$.

\subsection{Vertical settling}
\label{sec_vert_dr}

As it is assumed that dust streams only vertically, $g_x\to 0$, 
eq. (\ref{res_cond}) yields that IW and SDW couple to each other at
\begin{equation}
\label{res_cond_vert_appr}
k_x = (\tilde k_z^2 - k_z^2)^{1/2} \approx \tilde k_z
\end{equation}
in the long wavelength limit $k_z \ll \tilde k_z$. Hence, the long wave-length perturbations subject to RDI are extended along the vertical direction in the both cases of radially or vertically streaming dust. 
However, in the latter case, the shorter wave-length of 
perturbations along the radial direction $(k_x h)^{-1} \sim \tau$, while in the former case $(k_x h)^{-1} \sim \tau^{1/2}$. 
This difference almost eliminates the advantage of RDI of vertically streaming dust, which attains a greater growth rate independent of $\tau$, cf. eq. (\ref{gen_approx_sol}) and eq. (\ref{approx_sol_vert}) below.

Indeed, according to \citetalias{zhuravlev-2019} there is the following estimation for the maximum value of RDI growth rate 
for $g_x\to 0$ in the Keplerian disc
\begin{equation}
\label{approx_sol_vert}
\Delta \approx {\rm i}\, \Omega_0 \left ( \frac{f}{2} \right )^{1/2} \frac{k_x}{k}.
\end{equation}
Hence, in the limit given by eq. (\ref{res_cond_vert_appr}) 
eq. (\ref{approx_sol_vert}) yields with the account of dissipative effects
\begin{equation}
\label{Im_om_vert}
\frac{\Im[\omega]}{\Omega_0} \approx \left ( \frac{f}{2} \right )^{1/2} - \alpha h^2 \tilde k_z^2
\end{equation}
and the upper limit of $\alpha$ is
\begin{equation}
\label{alp_max_vert}
\alpha_{max} \approx \left ( \frac{f}{2} \right )^{1/2} \left ( \frac{z_0}{h} \right )^2 \tau^2. 
\end{equation}

Comparing equation (\ref{alp_max_vert}) with equation (\ref{alp_max_rad}) one finds that for small particles $\tau \ll 1$ 
the both variants of RDI are suppressed at $\alpha_{max} \propto f^{1/2} \tau^2$.

\subsection{Settling combined with radial drift}
\label{sec_comb_dr}

The situation considered in the previous Section changes as the small amount of the dust radial drift is taken into account.
Let perturbations be still extended along the vertical direction $k_z \ll k_x$ and 
\begin{equation}
\label{ineq_g_x_z}
g_z \gg g_x,
\end{equation}
but the vertical length-scale of perturbations to become so small that
\begin{equation}
\label{k_z_cond}
k_z \lesssim \frac{\tilde k_z^2}{\tilde k_x} \sim \left ( \frac{z_0}{h} \right )^{-1}  \frac{\eta}{\tau \delta_*} \, z_0^{-1}.
\end{equation}
Then, the first term in LHS of eq. (\ref{res_cond}) prevails the second term therein in spite of the condition (\ref{ineq_g_x_z}).
The mode crossing is located at $k_x$ given by eq. (\ref{res_cond_rad_appr}).
Eq. (\ref{k_z_cond}) shows 
that the regime of RDI considered in this Section is valid at $k_z \gtrsim z_0^{-1}$ 
for sufficiently small particles $\tau \sim \delta_*$. 
Since the dust streams both vertically and radially, RDI appears due to the type-I mode coupling of IW and SDW within TVA. 
The general expression for the highest RDI growth rate appropriate for this case
is adopted from \citetalias{zhuravlev-2019} for the Keplerian disc
\begin{equation}
\label{Delta_gen}
\Delta = {\rm i}\, \Omega_0 \left ( \frac{f}{2} \right )^{1/2} \frac{(k_x k_z)^{1/2}}{k} 
\left ( \frac{k_x g_z - k_z g_x}{g_x k_x + g_z k_z} \right )^{1/2}.
\end{equation}
In the just described wavenumber limit, it yields the following estimate for the growth rate with the account of dissipative effects
\footnote{The violation of (\ref{k_z_cond}) leads to the same estimate of the growth rate 
and $\alpha_{max}$ as was previously obtained for the case $g_x\to 0$, see Section \ref{sec_vert_dr}.}
\begin{equation}
\label{Im_om_rad_vert}
\frac{\Im[\omega]}{\Omega_0} = \left (\frac{f}{2} \right )^{1/2} \left ( \frac{k_z}{\tilde k_x} \right )^{1/4}
\left ( \frac{g_z}{g_x} \right )^{1/2} - \alpha h^2 k_z \tilde k_x 
\end{equation}
and 
\begin{equation}
\label{alp_max_rad_vert}
\alpha_{max} \approx f^{1/2} \left ( \frac{z_0}{h} \right )^{5/4} \tau^{5/4} \left [ \frac{\eta}{(k_z z_0) \delta_*} \right ]^{3/4}.
\end{equation}

Thus, as far as $z_0\sim h$, this regime of RDI operates up to the considerably higher effective viscosity as compared to 
the case of the dust settling alone, and more so, to the case of the dust drifting only radially. 
Physically, the presence of the dust settling allows for the existence of RDI in 
the zeroth order in $\tau$, while incorporation of the radial drift of the dust shifts the mode coupling to 
longer wavelengths $\sim \tau^{1/2}$ according to eq. (\ref{res_cond_rad_appr}). In this way, RDI persists for smaller 
particles and for higher value of the effective viscosity/diffusivity.
This result is demonstrated in Figs. \ref{fig_1} and \ref{fig_2}, where the pattern speed and the growth/damping rate of 
the mode of gas-dust perturbations are plotted vs. the radial wavenumber. In the four different cases the inviscid 
solution is compared with the damping solution. Each time the damping solution is shown for a particular value 
of the viscosity, which eliminates the growth rate of RDI at resonance between IW and SDW. 
As expected, for $Sc=1$ the solution of both of the dispersion equations, (\ref{disp}) and ({\ref{disp_next_order}}),  
is nothing but the inviscid solution shifted by the characteristic imaginary damping frequency. 
Analytical estimations performed in Sections \ref{sec_rad_dr}-\ref{sec_comb_dr}, see eqs. (\ref{Im_om_rad}), 
(\ref{Im_om_vert}) and (\ref{Im_om_rad_vert}), are in a good agreement with an exact solutions.

It is assumed that in Figs. \ref{fig_1} and \ref{fig_2} as well as in the rest of the Figures below,
the frequencies and wavenumbers are measured in units of 
$\Omega_0$ and $L^{-1}$, respectively, where $L\lesssim h$ is the artificial length-scale. Correspondingly, $g_{x,z}$ are 
measured in units of $\Omega_0^2 L$, while $\nu$ is measured in units of $\Omega_0 L^2$. 
Note that for the particular choice $L=h$, the value of $\nu$ equals to $\alpha$. 
Figs. \ref{fig_1} and \ref{fig_2} provide the mutual comparison of RDI in the 
three situations described analytically in Sections \ref{sec_rad_dr}-\ref{sec_comb_dr}. 
This is done for two particular values of $k_z$. 
At first, it is seen that RDI of the radially streaming dust is suppressed by a very small amount of dissipation 
even for formally the least value of $k_z h = 1$, which can be considered in the local approach. 
Of course, as one proceeds to the larger particles and the higher dust fraction, the threshold $\nu$ takes more reasonable values.
It can be checked that the threshold values of $\nu$ used to plot damping solutions in the both of the figures are in agreement with
the corresponding estimates provided by eqs. (\ref{alp_max_rad}), (\ref{alp_max_vert}) and (\ref{alp_max_rad_vert}). 
Thus, particularly for $k_z h = 1$ and for $f$, $\tau$ by an order of magnitude higher than used in Fig. \ref{fig_2} 
the threshold value of $\nu$ approaches $10^{-4}$. This is due to the higher growth rate along with the higher length-scale 
of RDI for larger particles. 

The above result changes as the dust settling is taken into account. For solids located sufficiently high above the disc midplane
the limit (\ref{ineq_g_x_z}) is normally true. Hence, for vertical wavenumber of perturbations higher than the value given by 
the restriction (\ref{k_z_cond}) RDI is suppressed according to the estimates, which are valid in neglecting the radial drift
of the dust, see Section \ref{sec_vert_dr}. For the particular $g_z=1.0$, $\tau=0.01$ taken in Fig. \ref{fig_1} this implies that 
the solution obtained for $k_z h = 40$ (or $k_z h = 1$) remains unaffected by the radial drift of the dust
as far as $g_x \lesssim 0.4$ (or $g_x \lesssim 0.01$).
In this limit, the corresponding threshold value of $\nu$ is independent of $k_z$, see eq. (\ref{alp_max_vert}). 
Therefore, as one looks at the lowest wavenumbers within the local approach of this study, $k_z h \sim 1$, RDI of the vertically
streaming dust is eliminated by dissipation at moderately higher threshold value of $\nu$ than that of RDI
of the radially drifting dust. However, the difference in the threshold values of $\nu$ in these two cases 
is not so large as might be expected from their typical growth rates in the inviscid limit, 
which differ by two orders of magnitude.
The reason for that is the radial wavenumber of the coupling modes given by eq. (\ref{res_cond_vert_appr}), 
which does not decrease as one goes to small $k_z$. 

However, as soon as $k_z$ becomes smaller than the restriction (\ref{k_z_cond}), the weak contribution of the radial drift 
into the dust settling modifies the picture of damping RDI once again. It turns out that IW and SDW coupling to each other 
shift back to the radial length-scale of RDI of the radially streaming dust, cf. eq (\ref{res_cond_rad_appr}) and 
eq. (\ref{res_cond_vert_appr}), which is longer as compared to the case of the dust settling alone. 
This considerably reduces dissipation. 
On the other hand, the corresponding RDI growth rate only weakly decreases with $k_z$, see eq. (\ref{Im_om_rad_vert}).
Consequently, one finds the threshold value of viscosity to become higher by more than an order of magnitude for the same 
$f=0.01$ and $\tau=0.01$, see the curves marked as `3' and `4' in Fig. \ref{fig_1}. 
Eq. (\ref{alp_max_rad_vert}) indicates that for $f$, $\tau$ by an order of magnitude higher than used in Fig. \ref{fig_1} 
the similar threshold value of $\nu$ exceeds $2\cdot 10^{-3}$, which is by more than an order of magnitude higher than
that of RDI of radially drifting dust, see the discussion above.
Moreover, as seen in Fig. \ref{fig_1}, in the last case RDI persists at the longer radial wavelengths comparing to the mode crossing.
This implies that strict stabilisation of gas-dust mixture occurs at even higher $\alpha > \alpha_{max}$.

\section{Viscous instability of the settling dust}
\label{sec_Sc_infty}

This Section is focused on the case when $Sc > 1$, which is commonly found in the numerical simulations of 
turbulence in protoplanetary discs.
As the Schmidt number differs from unity, $\omega_*$ is non-zero, so the solution of the original dispersion equation (\ref{disp}) can not be expressed via eq. (\ref{visc_invisc}). Instead, it is sought in the form of eq. (\ref{replace}), 
where $\omega^\prime$ obeys an eq. (\ref{disp_mod}). Eq. (\ref{disp_mod}) is no more reduced to the inviscid problem. 
Here this issue is addressed for the particular case of the dust settling, $g_x\to 0$, and considered within TVA.
Clearly, there is a degeneracy of the problem with respect to parameters $\nu$ and $Sc$. The solution 
is determined by their combination, which is $\nu_* \equiv \nu-D = \nu(1-Sc^{-1})$, see eq. (\ref{om_diff}).
As $Sc\gg 1$, then $\nu_* \to \nu$, while $D\to 0$, which implies that $\omega \to \omega^\prime$.
Oppositely, as $Sc\to 1$, $\nu_*\to 0$ for any constant $\nu$ and $D$, returning the solution back the form (\ref{visc_invisc}).
A behaviour of $\omega^\prime$ with respect to non-zero $\nu_*$ is discussed in this Section below.

In the absence of the dust backreaction on gas, i.e. as $f\to 0$, the coupling term in RHS of eq. (\ref{disp_mod}) vanishes, so 
there are three independent formal modes of $\omega^\prime$. Two of them are the oppositely propagating IW, which
formally decay at a rate of $\omega_*$, while the third mode is the formally neutral SDW, 
since it has a real auxiliary frequency, $\omega^\prime = \omega_p$.
Considering equation (\ref{disp}) in the inviscid limit, \citetalias{zhuravlev-2019} revealed that as 
there are SDW and IW, both of which are neutral, while SDW has negative energy, 
the non-zero coupling term causes their resonant coalescence into the coupled modes with equal growth/damping rates. 
This coalescence occurs in a finite interval of wavenumbers. 
Outside this interval the modes akin to SDW and IW remain neutral, however, their pattern speeds are modified 
by coupling.
Here, eq. (\ref{disp_mod})
describes how the coupling term modifies the negative energy SDW and the {\it damping} IW in the vicinity of their mode crossing. 
It is shown below, that such a process takes a quasi-resonant form leading to new instability.

\begin{figure}
\begin{center}
\includegraphics[width=8cm,angle=0]{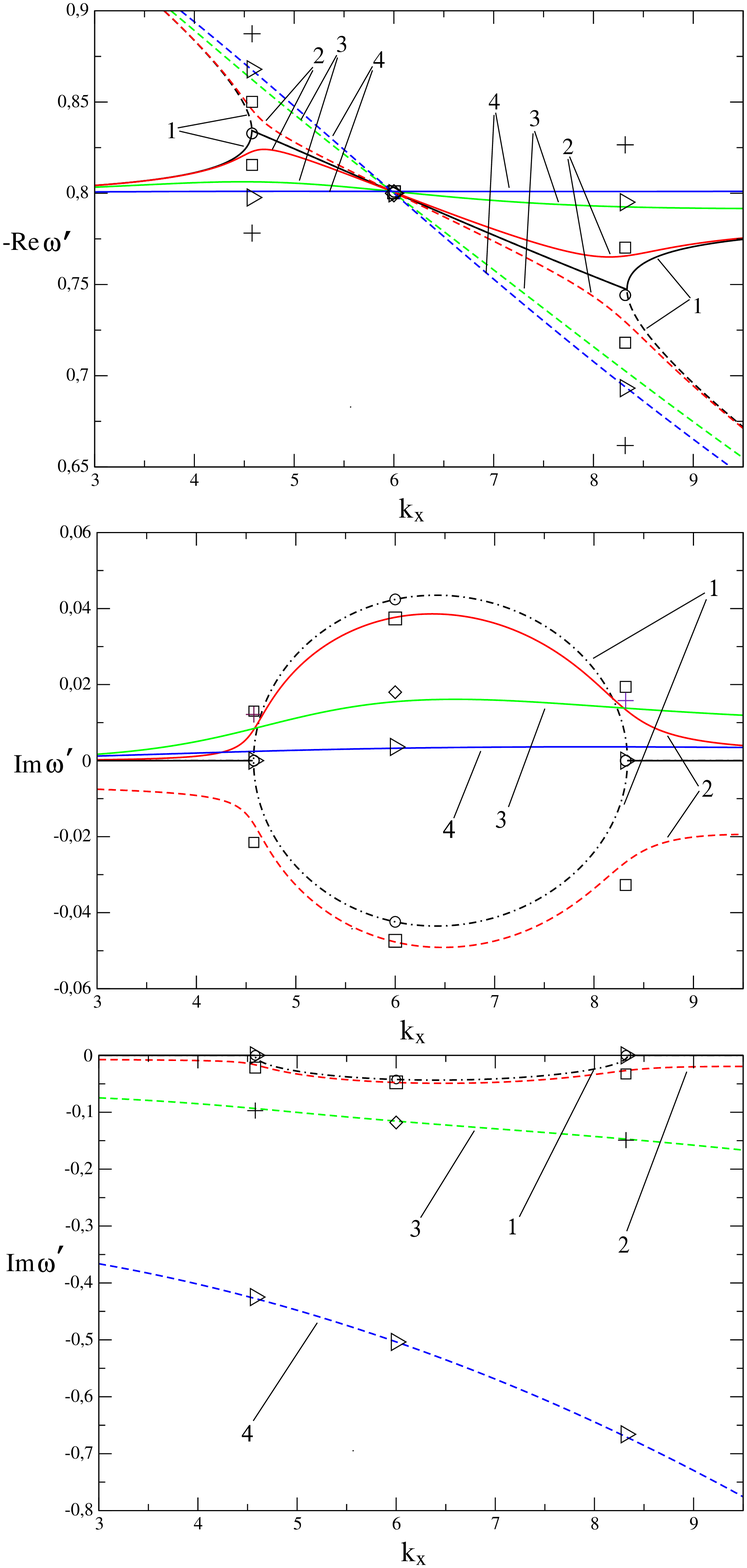}
\end{center}
\caption{ 
Top panel: the curves show $\Re[\omega^\prime]$ taken with the opposite sign,
where $\omega^\prime$ is the solution of the dispersion equation (\ref{disp_mod}). 
The parameters are $q = 3/2$, $f=0.01$, $g_x = 0.0$, $g_z = 1.0$, $\tau = 0.1$ and $k_z = 8$.
Solid and dashed lines represent the modes akin to SDW and IW, respectively, except for the solid line showing the 
band of instability for $\nu_*=0$, which represents the coupled modes. 
The curves marked as `1',`2',`3',`4' show the solution for, respectively, $\nu_* = 0, 0.0001, 0.001, 0.005$.
Middle and bottom panels: the imaginary part of the corresponding solution.
The notations are the same. Additionally, the dot-dashed line shows the coupled modes for $\nu_*=0$.
The analytical estimations: circles, squares and pluses represent the limit of weak viscosity, $c_{\nu} \ll f^{1/2}$ 
(see eqs. (\ref{approx_cross_weak_visc}), (\ref{approx_bounds_weak_visc_re}) and (\ref{approx_bounds_weak_visc_im}) ), 
for, respectively, $\nu_*=0, 0.0001, 0.001$; 
the diamonds and the triangles represent the limit of strong viscosity, $c_{\nu} \gg f^{1/2}$ 
(see eqs. (\ref{approx_cross_strong_visc}) and (\ref{approx_bounds_strong_visc}) ), for, respectively, $\nu_*=0.001$ and $\nu_*=0.005$. 
On the coloured versions of plots black, red, green and blue curves represent the solutions for, respectively, 
$\nu_* = 0, 0.0001, 0.001, 0.005$.
} 
\label{fig_3}
\end{figure}

An accurate numerical solution of eq. (\ref{disp_mod}) in the band of RDI for the particular values of $g_z$ and $\tau$ 
is presented in Fig. \ref{fig_3}. The vertical wavenumber is set to $k_z=8$, which places the mode crossing into $k_x=6$ 
in order to mimic the particular solution demonstrated in Fig. 1 of \citetalias{zhuravlev-2019} in the standard case $Sc=1$. 
The curves in Fig. \ref{fig_3} show $-\Re[\omega^\prime]$ and $\Im[\omega^\prime]$ for several values of $\nu_*$. 

For $\nu_*=0$ the curves are identical to the inviscid solution with respect to $\omega$, cf. Fig. 1 of \citetalias{zhuravlev-2019}. 
There is a complex conjugate pair of solutions having the identical pattern speeds $\Re[\omega^\prime]$, 
which represent the mode coupling of IW and SDW in the vicinity of the mode crossing. 
Outside the band of RDI the modes akin to IW and SDW acquire different pattern speeds, while losing their growth/decay. 
As soon as $\nu_*>0$, the modes become decoupled, i.e. they acquire different pattern speeds inside the band of RDI. 
Note that this is true throughout the whole band of RDI\footnote{Of course, except the mode crossing itself.} 
for any non-zero value of $\nu_*$. The difference of pattern speeds becomes higher as $\nu_*$ increases. 
Finally, $\Re[\omega^\prime]$ of both modes akin to IW and SDW approaches the values they would have in the absence of the mode coupling (for $f\to 0$), i.e. profiles of $\Re[\omega^\prime]$ for the original IW and SDW. 
This occurs for $\nu_*\gtrsim 0.005$.

As might be expected, decrement of damping mode and increment
of growing mode inside the band of RDI, respectively, increases and decreases as $\nu_*$ becomes larger.
A different situation occurs outside the band of RDI: the neutral modes transform into {\it growing} and damping modes. 
Therefore, the non-zero $\nu_*$ produces a new instability, which disappears as $\nu_*\to 0$. A comparison of panels in 
Fig. \ref{fig_3} allows one to find that this is the mode akin to SDW, which becomes growing in a larger range of $k_x$ 
comparing to the case of $\nu_*=0$. At the same time, the mode akin to IW decays at a rate higher than the damping rate of 
the original IW, $\omega_*$.
It can be seen, that the mode akin to SDW attains maximum growth rate for some small non-zero value of $\nu_*$. 
This growth rate ceases as $\nu_*$ increases further.
This suggests that the growth of mode akin to SDW (as considered with respect to $\omega^\prime$) 
is defined by two distinct physical mechanisms:\\ 
(i) the mode coupling, which is essentially inviscid in its nature; \\
(ii) the additional dissipation of the energy of both negative energy SDW and damping IW; this process 
has a quasi-resonant nature, since it occurs though the coupling term, however, the modes are decoupled from each other; 
the drain of negative energy causes the new growth of gas-dust perturbations.\\
Mechanisms (i) and (ii) are responsible for instability in, respectively, bounded and unbounded range of wavenumbers.
Mechanism (ii) might be referred to as viscous instability of the settling dust or 'settling viscous instability' (SVI).
It is similar to dissipative instability, for example, of the electron beam, see e.g. \citet{nezlin-1976}, or 
the flow waves on a vortex sheet, see e.g. \citet{cairns-1979}.

As soon as $Sc$ considerably exceeds unity, SVI may defer the suppression of RDI as the dissipation in a turbulent 
disc increases. In particular, $\alpha_{max}$ given by eq. (\ref{alp_max_vert}) and (\ref{alp_max_rad_vert}) for the case 
of $Sc=1$ will shift to the greater values for increasing $Sc>1$.
In the remaining part of the work, SVI is considered analytically.


\subsection{Approximate solution at the mode crossing}

The analytical solution of eq. (\ref{disp_mod}) at the mode crossing
\begin{equation}
\label{mode_crossing}
\omega^\prime_c \equiv \omega_i = \omega_p,
\end{equation}
which for the given $k_z$ takes place at $k_x$ defined by the first equality in eq. (\ref{res_cond_vert_appr}),
can be obtained expanding the Cardano solution of eq. (\ref{disp_mod}) by the orders of small $f^{1/2}$. 

A ratio
\begin{equation}
\label{c_nu}
c_\nu \equiv \frac{\omega_*}{\omega_i},
\end{equation}
can be used as dimensionless parameter describing the viscosity contribution into the analytical solution. 
It is assumed that $c_\nu$ is small, however, it can be in any ratio with $f^{1/2}$.


In the leading orders of $c_\nu$ and $f^{1/2}$ it is found that the dust back-reaction of gas along with dissipative effects in 
the gas-dust mixture produce the following corrections to $\omega^\prime_c$. 
In the limit of weak viscosity, $c_\nu \ll f^{1/2}$,
\begin{equation}
\label{approx_cross_weak_visc}
\frac{\omega^\prime_{\rm SDW,IW}}{|\omega^\prime_c|} \approx 1 -\frac{{\rm i} c_\nu}{2} \pm {\rm i} \beta \left ( \frac{f}{2} \right )^{1/2},
\end{equation}
while in the limit of strong viscosity, $c_\nu \gg f^{1/2}$, 
\begin{equation}
\label{approx_cross_strong_visc}
\frac{\omega^\prime_{\rm SDW, IW}}{|\omega^\prime_c|} \approx 1 + {\rm i} \frac{\beta^2}{2}\frac{f}{c_\nu},\quad
1 - {\rm i} c_\nu - {\rm i} \frac{\beta^2}{2}\frac{f}{c_\nu},
\end{equation}
where the dimensionless parameter
\begin{equation}
\label{beta}
\beta \equiv \frac{k_x}{k_z}
\end{equation}
along with $c_\nu$ are to be taken at the mode crossing. 
In eqs. (\ref{approx_cross_weak_visc}) and (\ref{approx_cross_strong_visc}) the subscripts 
`SDW' and `IW' mean that the corresponding frequency belongs to the mode akin to SDW or IW. 
It is seen that these analytical approximations 
are in a good agreement with the numerical values of $\omega^\prime$ for the corresponding modes in the both limits, 
see Fig. \ref{fig_3}.

It is important to note, that according to eq. (\ref{approx_cross_strong_visc}), the growth of SDW exists for any value 
of $\nu_*$, i.e. the neutral mode is achieved asymptotically, as $\nu_* \to \infty$.
This is a specific manifestation of SVI, since for $Sc=1$ and $\nu_*=0$, i.e. in the absence of formal damping of IW, 
the growing coupled mode is always stabilised by the finite threshold value of viscosity.

%
%

\subsection{Approximate solution at the bounds of the mode coupling}

It is instructive to show analytically the emergence of SVI in a broader band of wavenumbers as compared to the inviscid RDI.
For that, the bounds of the mode coupling can be chosen. The determinant of the Cardano solution of eq. (\ref{disp_mod}) 
for $\nu_*=0$ vanishes at these bounds, which is used here to find the corresponding wavenumbers. 
Expanding the Cardano solution of eq. (\ref{disp_mod}) by the orders of small $f^{1/2}$ at these bounds yields the approximate
frequencies for growing SDW and damping IW.
In the limit of weak viscosity, $c_\nu \ll f^{1/2}$, the following result is obtained:
\begin{equation}
\label{approx_bounds_weak_visc_re}
\frac{\Re[\omega^\prime_{\rm IW,SDW}]}{|\omega^\prime_c|} = 1 \pm \beta \left ( \frac{f}{2} \right )^{1/2} 
\left [ 1 \pm \left ( \frac{\beta (f/2)^{1/2}}{c_\nu} \right )^{-1/2} \right ],
\end{equation}
where the first $\pm$ defines the two bounds of the inviscid RDI, while the second $\pm$ inside the square brackets defines 
either SDW or IW. The growth/damping rate reads 
\begin{equation}
\label{approx_bounds_weak_visc_im}
\frac{\Im[\omega^\prime_{\rm IW,SDW}]}{|\omega^\prime_c|} = -\frac{c_\nu}{2} \left [ 1 \pm 2 \left ( \frac{\beta (f/2)^{1/2}}{c_\nu} \right )^{1/2} \right ].
\end{equation}

In the limit of strong viscosity, $c_\nu \gg f^{1/2}$, there is another result:
\begin{equation}
\label{approx_bounds_strong_visc}
\frac{\Re[\omega^\prime_{\rm SWD,IW}]}{|\omega^\prime_c|} = 1 \pm \beta (2 f)^{1/2},\quad  1 - {\rm i} c_\nu,
\end{equation}
where $\pm$ defines the two bounds of the inviscid RDI.

In eqs. (\ref{approx_bounds_weak_visc_re})-(\ref{approx_bounds_strong_visc}) the parameters $\beta$ and $c_\nu$
are to be evaluated at the corresponding bound of the inviscid RDI.
The straightforward analysis shows that for the given $k_z$ both of them are determined by the quadratic equation for $\beta$:
\begin{equation}
\label{eq_beta}
( 7 f - 2 \hat k_z^2 ) \beta^2 \pm 4 (2 f)^{1/2} \beta - 2 (\hat k_z^2 -1) = 0,
\end{equation}
where $\hat k_z \equiv k_z/\tilde k_z$.

\begin{figure}
\begin{center}
\includegraphics[width=8cm,angle=0]{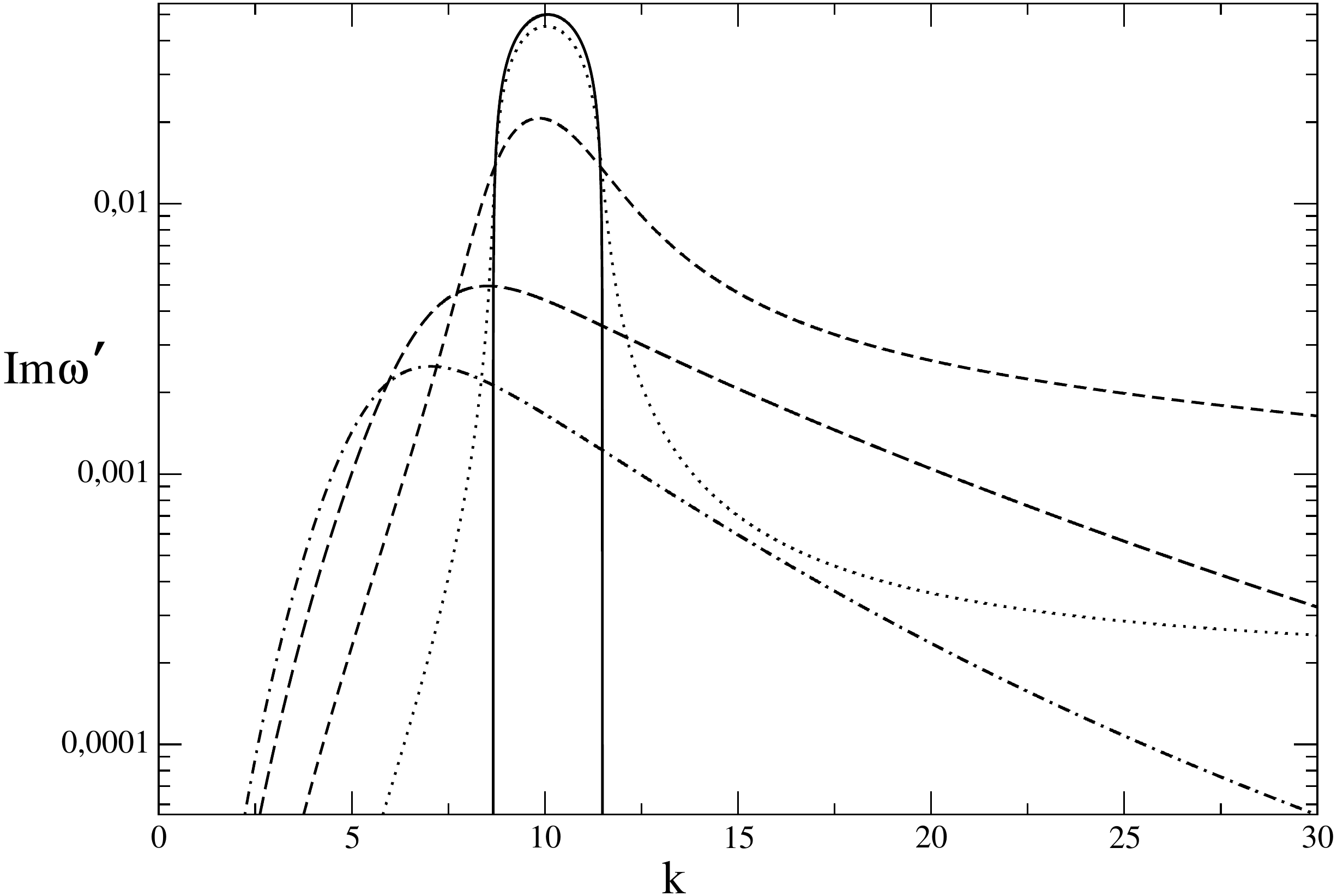}
\end{center}
\caption{Imaginary part of the solution of the dispersion equation (\ref{disp_mod}) corresponding to the mode akin to SDW 
vs. the absolute value of $k$ for $\theta=45^{\rm o}$.
Solid, dotted, short-dashed, long-dashed, dot-dashed curves represent, 
respectively, $\nu_* = 0, 0.0001, 0.001, 0.005, 0.01$.
Free parameters, notations and units are the same as in Fig. \ref{fig_3}.
}
\label{fig_4}
\end{figure}

The approximate solutions described in this Section are plotted in Fig. \ref{fig_3}. 
There is a satisfactory agreement with an accurate result, except for the application of the limit 
of weak viscosity to the case of $\nu_*=0.001$. The decoupling of $\Re[\omega^\prime]$ is overestimated according to the analytical
value, however, this might be not surprising as far as $c_\nu \sim f^{1/2}$ for $\nu_*=0.001$.
At the same time, eq. (\ref{approx_bounds_weak_visc_im}) obtained with the assumption $c_\nu \gg f^{1/2}$ reproduces well an accurate value of SVI growth rate up to $c_\nu \sim f^{1/2}$. The analytical growth rate attains maximum at $c_\nu=\beta (f/2)^{1/2}$, 
which is explicitly
\begin{equation}
\label{max_Im_SVI}
\max \frac{\Im[\omega^\prime_{\rm SDW}]}{|\omega^\prime_c|} = \frac{\beta}{2} \left ( \frac{f}{2} \right )^{1/2}.
\end{equation}
Clearly, eq. (\ref{max_Im_SVI}) provides even higher values as compared to the inviscid RDI growth rate, 
see eq. (\ref{approx_sol_vert}). This result underlines the importance of SVI.


The substantial growth rate of SVI is attained at the formal bound of the regime of weak viscosity, $c_\nu \sim \beta f^{1/2}$.
It is instructive to check which value of $\alpha$ this corresponds to. As $\beta \gg 1$ and the flow is Keplerian, eq. (\ref{c_nu})
yields
\begin{equation}
\label{alp_max_SVI}
\alpha_{\rm SVI} \sim \frac{Sc}{Sc-1} \left ( \frac{f}{2} \right )^{1/2} \left ( \frac{z_0}{h} \right )^{2} \tau^2.
\end{equation}
For $Sc$ considerably higher than unity $\alpha_{\rm SVI}$ approaches $\alpha_{max}$ provided by eq. (\ref{alp_max_vert}), 
but never goes below this value. This implies that the level of the effective viscosity sufficient to suppress RDI 
for $Sc=1$ also corresponds to the highest growth rate of SVI given by eq. (\ref{max_Im_SVI}).
Therefore, SVI always enlarges the domain of RDI in turbulent discs with $Sc>1$. Additional numerical tests indicate that 
$\alpha_{max}$ increases at least by factor of $\sim 2 $ as $Sc\to 5$. 
Besides, SVI expands the band of damping RDI corresponding to the case $Sc=1$ to both small and large wavenumbers, 
see Fig. \ref{fig_4}.

\begin{figure}
\begin{center}
\includegraphics[width=8cm,angle=0]{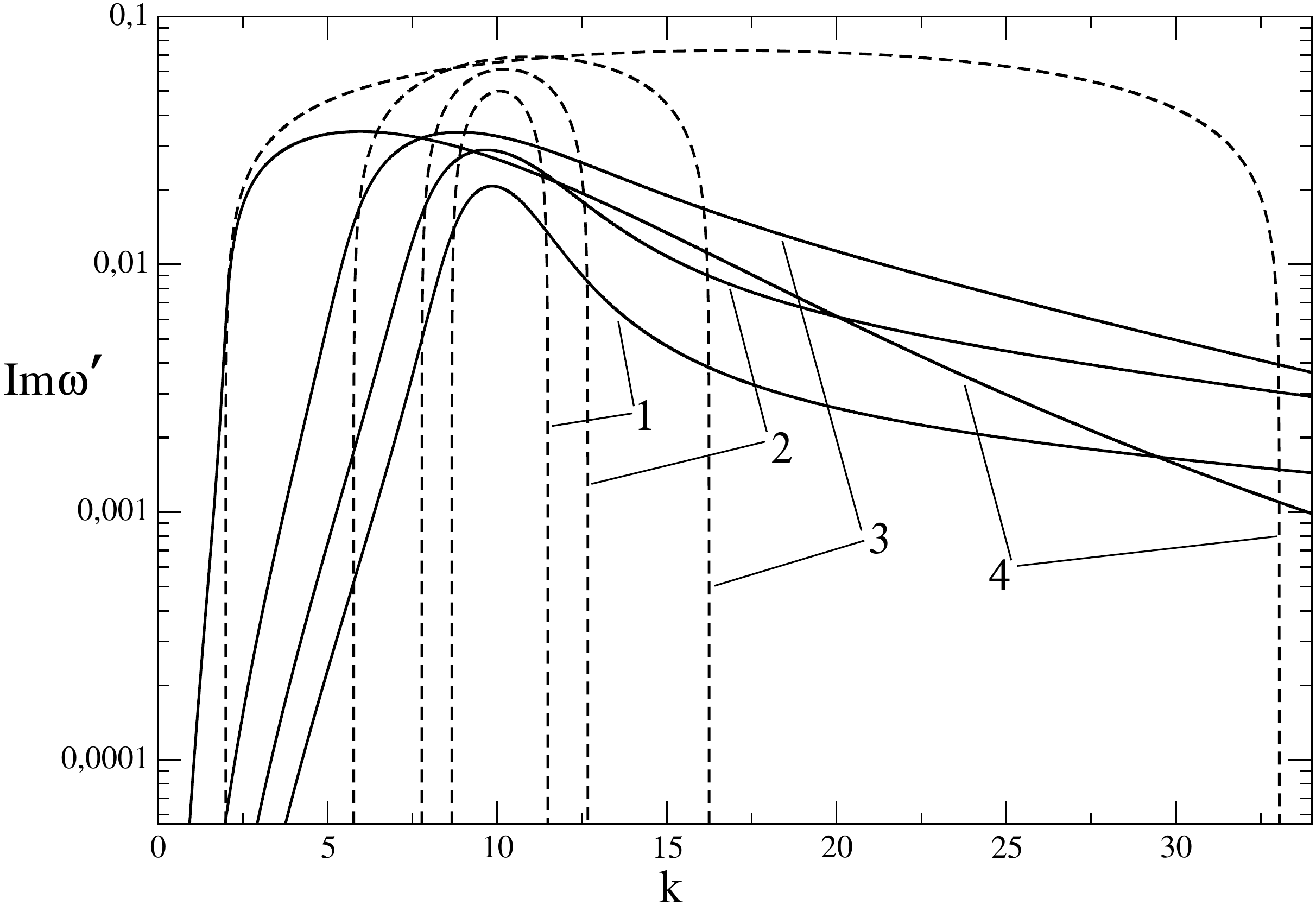}
\end{center}
\caption{The same as in fig. \ref{fig_4} but for various $\theta=45^o, 60^o, 75^o, 85^o$ corresponding to curves marked as '1', '2', 
'3', '4'. The viscosity is $\nu_*=0.001$. The dashed curves show the corresponding inviscid solution. 
} \label{fig_5}
\end{figure}

\subsection{Approximate solution for small wavenumbers}

Figs. \ref{fig_4} and \ref{fig_5} show SVI in the same situation as in Fig. \ref{fig_3} but for a broader range of wavenumbers.
The value of $\beta$ is set to constant through the angle of the mode direction, $\theta$, which is
$$
\theta \equiv {\rm arctg}\, \beta.
$$

The behaviour of SVI in the limit of small $k$ is of the main interest here.
It can be seen that at the side of small $k$ the SVI growth rate {\it increases monotonically} as $\nu_*$ becomes larger.
It increases also for larger $\beta$. 
This feature of SVI may be especially important as a physical application to evolution of dust in protoplanetary discs 
bearing in mind that while $k\to 0$ the damping 
correction in eq. (\ref{replace}) vanishes. 
Whether this correction is stronger of weaker than SVI growth rate, can be clarified through the analytical approach 
to the solution of eq. (\ref{disp_mod}). 

First, eq. (\ref{disp_mod}) is written in the dimensionless form
\begin{equation}
\label{disp_small_k}
\hat D^\prime_g \hat D^\prime_p = -f \hat\omega_p \beta^2,
\end{equation}
where $\hat D^\prime_g \equiv (\hat \omega^\prime + {\rm i}\hat\omega_*)^2-1$, 
$\hat D^\prime_p \equiv \hat \omega^\prime-\hat\omega_p$, 
while
$\hat\omega^\prime \equiv \omega^\prime/\omega_i$, 
$\hat \omega_p \equiv \omega_p/\omega_i$,
$\hat\omega_* \equiv \omega_*/\omega_i$.

Eq. (\ref{disp_small_k}) is considered in the limit of $k\ll \tilde k_z$ and $\beta = const$.
Therefore, the frequency of IW, $\omega_i$, remains constant, while the frequency of SDW 
becomes the small value, $\hat \omega_p \ll 1$. Note that in the same limit $\hat\omega_*$ may still be a value comparable to unity
or even higher than unity.

\begin{figure}
\begin{center}
\includegraphics[width=8cm,angle=0]{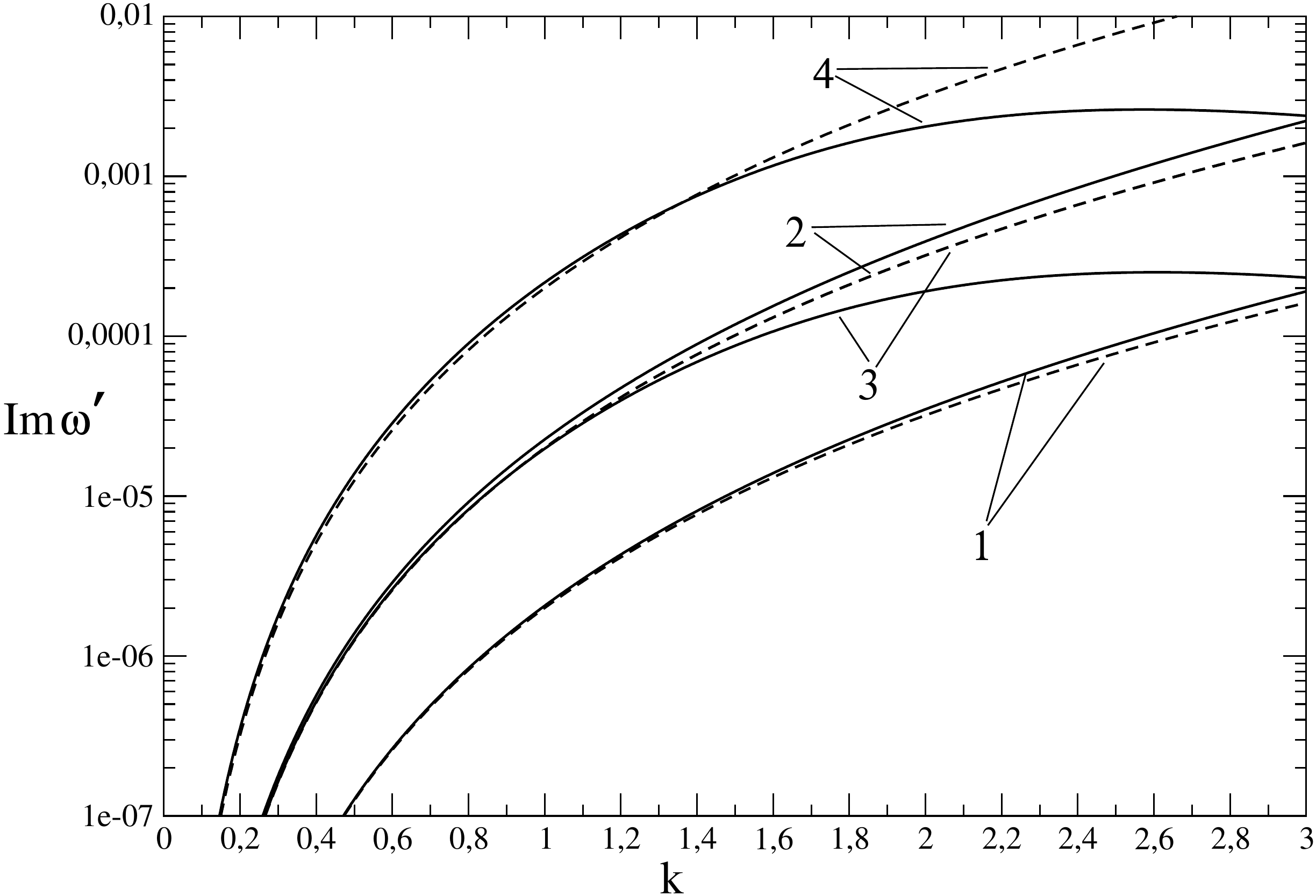}
\end{center}
\caption{Imaginary part of the solution of dispersion equation (\ref{disp_mod}) vs. the absolute value of $k$ for $k_x=k_z$.
Parameters, notations and units are the same as in Fig. \ref{fig_3}.
Solid and dashed curves represent, respectively, the solution of dispersion equation and the analytical solution in the limit of
the small wavenumber given by eq. (\ref{om_small_k}). 
Curves marked by '1', '2', '3' and '4' correspond to values of $f,\nu_*$ as follows:
$(0.01, 0.01)$, $(0.1, 0.01)$, $(0.01, 0.1)$ and $(0.1, 0.1)$.
} \label{fig_6}
\end{figure}

Further, as the dust fraction vanishes, there is an exact solution describing SDW:
\begin{equation}
\label{SDW_small_k}
\hat \omega^\prime = \hat \omega_p.
\end{equation}
Note that the mode coupling occurs at $\hat \omega^\prime = 1$.
For the small non-zero dust fraction the solution of eq. (\ref{disp_small_k}) slightly deviates from (\ref{SDW_small_k}):
$$
\hat \omega^\prime = \hat \omega_p + \Delta \hat \omega^\prime.
$$
Expansion of $\hat D^\prime_{g,p}$ up to the linear order in the small $\Delta \hat \omega^\prime$ in the vicinity 
of $\hat \omega_p$ replaces eq. (\ref{disp_small_k}) by the quadratic equation with respect to $\Delta \hat \omega^\prime$, 
which yields an approximate solution in the small wavenumber limit:
\begin{equation}
\label{om_small_k}
\hat \omega^\prime \approx \hat \omega_p + \frac{ f \beta^2 \hat \omega_p }{ 1 - (\hat \omega_p + {\rm i}\hat \omega_*)^2 }.
\end{equation}
It was checked that the imaginary part of the solution given by eq. (\ref{om_small_k}) is in a good accordance with an accurate 
solution of eq. (\ref{disp_mod}) in the small wavenumber limit, particularly for $\hat \omega_* \ll 1$, see Fig. \ref{fig_6}.
As far as $\hat \omega_* \ll 1$, both the imaginary part of eq. (\ref{om_small_k}) and the damping correction in eq. (\ref{replace})
are proportional to $\hat \omega_*$. 
Changing back to the dimensional variables, one obtains the approximate growth rate of SVI in this case,
\begin{equation}
\label{Im_om_small_k}
\Im[\omega] \biggr |_{\hat\omega_*\ll 1} \approx 
\frac{\omega_\nu}{Sc} \left [ 2 f \frac{k_x^2}{k_z^2} \frac{k^2}{\tilde k_z^2} (Sc-1) - 1 \right ],
\end{equation}
with an accurate dependence upon $Sc$. In eq. (\ref{Im_om_small_k}), $k_x$ and $k_z$ can be in any ratio with each other, while 
it is implied that $k\ll \tilde k_z$. Note that $k=\tilde k_z$ corresponds to the mode crossing. 
Eq. (\ref{Im_om_small_k}) does not provide a threshold value of viscosity as is the case for damping of RDI at $Sc=1$.
Instead, the small wavenumber SVI is proportional to the effective viscosity. It exists 
provided that combination in the square brackets is positive. The latter is true in the limit $k_z \ll k_x$, $k\approx k_x$, 
when
\begin{equation}
\label{SVI_cond_k}
\tilde k_z \gg k \gtrsim (\tilde k_z k_z)^{1/2} f^{-1/4} (Sc-1)^{-1/4}.
\end{equation}

On the other hand, eq. (\ref{Im_om_small_k}) is valid as far as
\begin{equation}
\label{om_*_restr}
\hat \omega_* = \alpha (1-Sc^{-1}) \frac{k}{k_z} \frac{k^2}{\tilde k_z^2} 
\left ( \frac{z_0}{h} \tau \right )^{-2} \lesssim 1 
\end{equation}
constructed with the help of eqs. (\ref{visc_alpha}), (\ref{g_z}), (\ref{om_nu}), (\ref{om_diff}) and (\ref{t_k_x}).
According to eq. (\ref{om_small_k}), SVI growth rate vanishes as $\sim 1/\hat \omega_*^3$ 
in the limit $\hat \omega_*\gg 1$.
Thus, the bound of SVI can be estimated using the marginal condition $\hat \omega_* \simeq 1$ from eq. (\ref{om_*_restr}).
That is, at a given effective viscosity eq. (\ref{om_*_restr}) puts the upper limit on the value of $k$ when SVI still 
operates. In this way, the lowest $k$ in eq. (\ref{SVI_cond_k}) yields the upper limit of effective viscosity 
for the existence of SVI:
\begin{equation}
\label{alp_max_SVI}
\alpha_{max} = f^{3/4} \tau^{3/2} \frac{Sc}{(Sc-1)^{1/4}} \left ( \frac{z_0}{h} \right )^2 (k_z z_0)^{-1/2},
\end{equation}
where it is employed eq. (\ref{t_k_z_h}).

Estimation (\ref{alp_max_SVI}) should be used in combination with
the requirement that there exists a band of $k$ identified in eq. (\ref{SVI_cond_k}), which reads
\begin{equation}
\label{cond_SVI_simple}
\tau \ll f^{1/2} (Sc-1)^{1/2} (k_z z_0)^{-1}.
\end{equation}
Apparently, eq. (\ref{cond_SVI_simple}) guaranteers the correct behaviour of $\alpha_{max}$ given by eq. (\ref{alp_max_SVI})
in the limit $Sc\to 1$: $\alpha_{max}\to 0$.

Eq. (\ref{Im_om_small_k}) taken for $\hat \omega_* \simeq 1$ shows that the low wavenumber SVI 
should attain an appreciably high growth rate of the order of $\sim (f/\tau) \Omega_0$ as $k\gg k_z$ approaches 
$\tilde k_z$ and the marginal case of perturbations with $k_z z_0 \sim 1$ is considered.
At the same time, SVI extends far into the region of low wavenumbers as can be found from eq. (\ref{SVI_cond_k}) 
for perturbations with $k_z z_0 \sim 1$. The latter can be important for the formation of even heavier dust clumps 
than in the case of RDI of the dust settling in laminar disc.

As soon as $Sc$ is considerably higher than unity, the necessary condition for the existence of the 
low wavenumber SVI given by eq. (\ref{cond_SVI_simple}) with $k_z z_0 \sim 1$
is definitely valid for newborn discs, where typically $\tau \ll f$. At the same time, it should be valid also for the
later stages of disc evolution, when dust particles grow up and approach the disc midplane, so that 
$\tau \sim f$ or even $\tau \gtrsim f$. 

Note that the threshold value of viscosity depends on $\tau$ and $f$ slightly steeper than in the case
of $Sc=1$ for settling combined with a small amount of radial drift of the dust, cf. eq. (\ref{alp_max_rad_vert}) and 
eq. (\ref{alp_max_SVI}). However, there is no dependence on $\eta/\delta_*$ in eq. (\ref{alp_max_SVI}), which implies that 
the low wavelength SVI persists up to the highest values of effective viscosity in the sufficiently thin discs or/and
at the sufficiently late stages of dust evolution, when $\tau$ and $f$ become closer to unity.

\subsection{The case $Sc<1$.}

The numerical solution of eq. (\ref{disp}) shows that the exchange 
$\omega_\nu \rightleftarrows \omega_D$ leads to the exchange of $\Re[\omega]$ between the roots representing the modes 
akin to SDW and IW, which produce damping RDI and SVI. 
At the same time, their imaginary parts remain the same. This result is confirmed by 
the analytical consideration of eq. (\ref{disp}) following \citetalias{zhuravlev-2019}, see his equations (72-74). 
The corresponding pair of approximate roots of eq. (\ref{disp}) obtained in the vicinity of the mode crossing is invariant with respect to $\omega_\nu \rightleftarrows \omega_D$.

Thereby, as $D > \nu$, SVI is produced by the growing mode akin to IW rather than SDW. For $Sc<1$ the new replacement 
$\omega \equiv \omega^{\prime \prime} - {\rm i}\omega_\nu$ reduces the problem to the coupling of formally {\it damping } SDW 
with neutral positive energy IW. As far as SDW is negative energy mode, its damping can only be caused by the {\it negative} dissipation. Clearly, the negative dissipation in the system produces the energy rather than takes it away.
Following an explanation given above for $Sc>1$, the coupling of modes in the case $Sc<1$ leads to an additional production 
of energy, which causes an extra damping of SDW and the growth of IW.

\section{Summary}

The linear instability of dust settling to the midplane of protoplanetary disc is a promising mechanism for clumping of solids
in particular disc ring providing their direct collapse into planetesimals or promoting the streaming instability 
of thin dust layer at the later stage of planetesimal formation. Its resistance against turbulence inherent in discs is 
an important issue to be resolved. This work introduces the first results obtained on that way by the analytical means.

Following \citetalias{zhuravlev-2019}, the problem is considered in the small shearing box approximation. It is assumed that 
length-scale of gas-dust perturbations is smaller than the disc scaleheight, which implies that the flow of gas is 
divergence-free. The analysis of the general equations for the centre-of-mass velocity of gas-dust mixture 
and the relative velocity of
gas and dust shows that in the usual limit of the particle stopping time and the particle stopping length much smaller
than the Keplerian time and length-scale of perturbations, respectively, $\tau\ll 1$ and $\lambda^{-1} \ll 1$, the dissipative terms
are not necessarily small, as they are ordered by the specific dimensionless parameter $\tau_\nu$ defined by eq. (\ref{tau_nu})
and limited according to eq. (\ref{tau_nu_est}). It is difficult to consider the general problem for $\tau_\nu\sim 1$ analytically,
see the Appendix for details. However, the resonance condition for RDI of the settling dust as given by eq. (\ref{l_ev_RDI})
along with the restriction for the bulk settling of dust in turbulent disc given by eq. (\ref{restr_stat_1}),
gives that $\tau_\nu$ is of the order of unity in the marginal case $\tau \sim \alpha$ corresponding to termination 
of the bulk settling across the disc, while it becomes small, $\tau_\nu\ll 1$, as far as the level of turbulence decreases and 
the smaller height $z_0<h$ is allowed for the bulk settling of dust. 
It can be checked that the similar conclusion is drawn about the RDI of the radially drifting dust.
Therefore, the assumption of $\tau_\nu\ll 1$ is justified as long as the problem of interest is the damping of RDI.

The restriction on the level of turbulence, which does not prevent the bulk settling of particles of the particular size, 
raises a concern whether turbulence smears the settling dust before it considerably affects the growing 
modes of gas-dust perturbations, as $\alpha$ increases. If not, the model used in this study is physically meaningful. 
The condition (\ref{restr_stat_6}) guaranteers that damping of RDI becomes substantial within the 
range of $\alpha$ allowed for the bulk settling of dust to coexist with turbulent fluctuations. Therefore, 
the smallness of $\alpha\ll 1$ makes the model used here be applicable to real discs. 
Note that as the bulk settling is inhibited by turbulence, the stationary solution (\ref{bg_U}-\ref{bg_sigma}) should be 
replaced by a different one taking into account gradient of the dust density. This option may be addressed in future along with 
the effects coming from gradient of the gas density, see eq. (\ref{term_diff}).

The assumption of $\tau_\nu \ll 1$ greatly simplifies the derivation of the dispersion equation for dissipative gas-dust 
modes of perturbations. In order to study RDI, the problem is considered in the leading order in the small
dust fraction, $f \ll 1$.
The dispersion equations for RDI have been derived in the two cases:\\
i) dust streaming both vertically and radially 
as considered in the leading order in $\tau \ll 1$, see eq. (\ref{disp}). \\
ii) the radial drift of the dust as considered in the next order in $\tau \ll 1$
comparing to (i), see eq. (\ref{disp_next_order}). \\

An additional assumption of viscosity and diffusivity equal to each other, $Sc=1$, 
reduces the solution of both eq. (\ref{disp}) and eq. (\ref{disp_next_order}) to the corresponding inviscid solution. 
This means that pattern speed of the mode remains unaffected by 
turbulent dissipation, while the imaginary part of the mode frequency is shifted by 
the characteristic damping frequency, $\omega_\nu$, 
see eq. (\ref{visc_invisc}). This is followed by the three estimations of RDI maximum growth rate: for 
the streaming instability (ii), see eq. (\ref{Im_om_rad}), 
for the settling instability of the dust streaming vertically (i), see eq. (\ref{Im_om_vert}), and 
for the settling instability of the dust streaming {\it almost} vertically (i), see eq. (\ref{Im_om_rad_vert}).
Thus, RDI ceases to operate at the threshold viscosity, which takes the three corresponding values, 
see eqs. (\ref{alp_max_rad}), (\ref{alp_max_vert}) and (\ref{alp_max_rad_vert}). 
That is combination of settling and small amount of the radial drift of the dust that provides the best persistence 
of the settling instability with respect to 
turbulent dissipation. Foremost, the dust settling provides the high growth rate independent of $\tau$ in contrast to 
what is the case for the streaming instability. However, the advantage in the growth rate leads to a moderate increase
of the threshold $\alpha=\alpha_{max}$, cf. eq (\ref{alp_max_rad}) and (\ref{alp_max_vert}), since RDI of vertically settling dust
does not have a long wave-length asymptotic, see eq. (\ref{res_cond_vert_appr}) for the resonance condition.
Nevertheless, the small amount of the dust radial drift in addition to the dust settling 
makes it possible for the band of RDI to shift to longer wave-lengths, as IW and SDW propagate 
almost radially, $k_x \gg k_z$, thus, reducing dissipation. This is illustrated in Fig. \ref{fig_1}. 
That is what gives a significant advantage to the settling instability in comparison with the streaming instability 
in a turbulent disc: the threshold viscosity of the former is substantially higher, compare Figs. \ref{fig_1} and \ref{fig_2}.

A novel instability arises in a turbulent gas-dust mixture with the bulk settling of dust as soon as diffusivity 
deviates from viscosity. The problem is no more reduced to the inviscid one being the coupling of SDW and IW 
which are both formally neutral. 
Instead, when $Sc>1$, it can be reduced to the coupling of the formally neutral SDW with {\it damping} IW. 
In this work, this case is considered in detail both numerically and analytically, see Fig. \ref{fig_3} and eq. (\ref{max_Im_SVI}).
Otherwise, when $Sc<1$, the problem is reduced to the coupling of the formally neutral IW with {\it damping} SDW. 
In the both cases, the effective damping frequency equals to the absolute value of the difference between 
the damping rates of the original SDW and IW in the absence of dust backreaction on gas, see eq. (\ref{om_diff}).
While for $Sc>1$ the damping IW introduces the true dissipation of energy, for $Sc<1$ the damping SDW 
introduces the {\it negative} dissipation, in other words, the production of energy. This is because SDW is the negative 
energy wave. 
Hence, in contrast to what occurs to the positive energy wave, 
its amplitude is decreased by the addition of the energy.
In this situation, the coupling between SDW and IW makes the mode akin to SDW (IW) growing in the unbounded range of wavelengths around the mode crossing for 
$Sc>1$ ($Sc<1$). The corresponding growth rate attains maximum in the vicinity of the mode crossing, 
while gradually decreasing as moving away to either side of the mode crossing. 
At the same time, the coupling makes the mode akin to IW (SDW) more damping for $Sc>1$ ($Sc<1$) as compared to the inviscid solution. 
Modification of both SDW and IW for both $Sc>1$ and $Sc<1$ can be thought of as quasi-resonant 
process, since it does not lead to coalescence of modes, as this is the case for the inviscid RDI. Instead, the modes 
save slightly different pattern speeds. The similarity to problems of the mode coupling in plasma physics and 
hydrodynamics of shear flows
allows one to borrow interpretation for the new instability. Specifically, the coupling of modes in dissipative system leads to 
an additional loss (production) of energy for both modes for $Sc>1$ ($Sc<1$) comparing to the inviscid coupling. 
This, in turn, leads to the growth of SDW (IW) when $Sc>1$ ($Sc<1$) since its energy is negative (positive). 
The new instability is referred to as SVI in this study. 
As follows from the description given above, the negative energy of SDW is the fundamental reason for SVI at any $Sc\neq 1$.
Besides, an explanation of SVI gives a clue why the solution of eq. (\ref{disp}) 
is reduced to the inviscid one for $Sc=1$, see eq. (\ref{visc_invisc}).
In the latter symmetric case, the true dissipation is precisely balanced by the negative dissipation in the system 
of the coupling IW and SDW.

At first, SVI manifests itself as the strengthening of RDI, see the discussion of estimations (\ref{max_Im_SVI})
in the text. 
Furthermore, SVI produces a long wavelength branch of instability, which vanishes as soon as $Sc\to 1$.
It is revealed that for sufficiently weak effective viscosity the growth rate of the long wavelength SVI is 
proportional to $\nu$, see eq. (\ref{Im_om_small_k}) derived for $Sc>1$.
This is confirmed by the accurate solution of the dispersion equation, see Fig. \ref{fig_4} (and also Fig. \ref{fig_6}), 
where it is seen that the growth rate
of SVI decreases close to the band of the inviscid RDI and increases at the small wavenumbers as the viscosity becomes larger.
The long wavelength SVI becomes stronger for almost radially propagating waves of gas-dust perturbations, $k_x \gg k_z$, which
is similar to the inviscid RDI of the settling dust, see Fig. \ref{fig_5}. In the marginal case of modes with 
the vertical length-scale approaching the dusty disc scale-height, the condition for the existence of the long wavelength SVI looks
physically plausible, see eq. (\ref{cond_SVI_simple}).

The analytical results obtained in this work, especially for $Sc\neq 1$, 
favour the instability of the dust settling to the disc midplane 
as an early mechanism for generation of the dust overdensities in the long-standing problem of planetesimal formation. 
Whether the ability of this instability to persist in the presence of the intrinsic shear turbulence 
remains on the non-linear stage of the growth of dust density perturbations is an issue for future research. 
On the other side, much simulation work on the instability of dust streaming in a (turbulent) protoplanetary disc 
remains to be done. The existing studies were focused on the instability of the radially drifting solids, 
which make up thin and dense dust layer on the late stage of dust coagulation in a disc. 
Also, turbulence commonly originated due to the magnetorotational instability. 
It is currently unclear how the number of suggested hydrodynamical mechanisms, 
see e.g. \citet{marcus-2015}, \citet{klahr-hubbard-2014} and \citet{barker-latter-2015}, which may be responsible 
for turbulence of another origin, affects RDI of the dust streaming either radially (the streaming instability), 
or vertically (the settling instability). Additionally, the shear turbulence in a disc may well be anisotropic, which makes 
the effective viscosity and diffusivity not only differ from each other but vary with the direction in a disc, 
see \citet{zhu-stone-bai-2015}.

\section*{Acknowledgments}

The author acknowledges the support from the Program of development of M.V. Lomonosov Moscow State University (Leading Scientific School 'Physics of stars, relativistic objects and galaxies').

\newpage


\appendix

%
%
%
%
%
%
%
%
%
%
%

\section{Description of main variables}

\label{app_symb}

\begin{tabular}{ll}
Symbol     & Meaning    \\ 
\hline
$\rho_g$  &  gas volume density  \\

$\rho_p$  &  dust volume density \\

$\rho$ & total density of gas-dust mixture \\

$f$ & dust fraction \\

$p$  & gas pressure \\

${\bf U}_g$ & velocity of gas \\

${\bf U}_p$ & velocity of dust \\

${\bf U}$ & centre-of-mass velocity of gas-dust mixture \\ 

${\bf V}$ & relative velocity of gas-dust mixture \\

${\bf u}$ & the Eulerian perturbation of ${\bf U}$ \\

${\bf v}$ & the Eulerian perturbation of ${\bf V}$ \\

$\rho_p^\prime$ & the Eulerian perturbation of $\rho_p$ \\

$p^\prime$ & the Eulerian perturbation of $p$ \\

$\delta$ & the relative perturbation of $\rho_p$ \\

$\nu$ & kinematic viscosity of gas \\

$D$ & dust diffusion coefficient \\

$\nu_*$ & the difference between $\nu$ and $D$ \\

$Sc$ & Schmidt number \\

$\Omega_0$ & local angular velocity of disc \\

$q$ & local disc shear rate \\

$\kappa$ & epicyclic frequency \\ 

$t_s$ & particle stopping time \\

$l_s$ & particle stopping length \\

$\tau$ & the Stokes number \\

$z_0$ & local height above the disc midplane \\

$h$ & disc scaleheight \\

$\delta_*$ & disc aspect ratio \\

$\alpha$ & Shakura-Sunyaev viscosity parameter \\

$k_x$ & radial wavenumber of mode \\

$k_z$ & vertical wavenumber of mode \\

$k$ & absolute value of wavenumber \\

$\omega$ & the mode frequency \\

$g_x$ & radial component of the effective stellar gravity \\

$g_z$ & vertical component of the effective stellar gravity \\

$\omega_\nu$ & characteristic frequency of viscous damping \\

$\omega_D$ & characteristic frequency of diffusive damping \\

$\hat \chi$ & the mode Fourier amplitude of any perturbation \\ 

$\omega_i$ & frequency of inertial wave \\

$\omega_p$ & frequency of the streaming dust wave \\

$\omega_*$ & the difference between $\omega_\nu$ and $\omega_D$ \\

$\omega^\prime$ & the mode frequency corrected for diffusive damping \\

$\alpha_{max}$ & the threshold viscosity of RDI or SVI \\

$\Delta$ & estimate of the maximum inviscid growth rate of RDI \\

$c_\nu$ & ratio of $\omega_*$ and $\omega_i$ \\

$\beta$ & ratio of $k_x$ and $k_z$ \\

$D_g$ & dispersion relation for inertial waves \\

$D_p$ & dispersion relation for the streaming dust wave \\

$\epsilon$ & coupling term for modes within TVA \\

$\tilde \epsilon$ & coupling term for modes beyond TVA \\

\hline
\end{tabular}

\newpage

\section{The general equations for small perturbations considered in TVA}

\label{app_pert}

The set of equations for non-axisymmetric perturbations in the leading order in $\tau$ is derived here with full account for 
the finite dust fraction.
The linearised eqs. (\ref{eq_U_2}), ({\ref{eq_TVA}), (\ref{eq_rho}), (\ref{eq_rho_tot}) read 

\begin{equation}
\label{pert_U}
\begin{aligned}
{\cal D}\, {\bf u} - 2\Omega_0 u_y {\bf e}_x + (2-q)\Omega_0 u_x {\bf e}_y + \\ 
({\bf u}\nabla){\bf U} + ({\bf U}\nabla) {\bf u} =  
-\frac{\nabla p^\prime}{\rho} + \frac{\nabla (p+p_0)}{\rho} \frac{\rho_p^\prime}{\rho} + {\bf F}_\nu,
\end{aligned}
\end{equation}

\begin{equation}
\label{pert_V}
\frac{\nabla p^\prime}{\rho} - \frac{\nabla(p+p_0)}{\rho} \frac{\rho_p^\prime}{\rho} = \frac{{\bf v}}{t_s} + {\bf G}_\nu,
\end{equation}

\begin{equation}
\label{pert_rho}
\nabla \cdot {\bf u}_g = 0,
\end{equation}

\begin{equation}
\label{pert_rho_tot}
{\cal D}\, \rho_p^\prime + \nabla (\rho_p^\prime {\bf U} + \rho\, {\bf u})  = 
D \nabla^2 \rho_p^{\prime},
\end{equation}
where the dissipative terms are explicitly

\begin{equation}
\label{F_nu} 
\begin{aligned}
{\bf F}_\nu = -\frac{\rho_g}{\rho} \frac{\rho_p^\prime}{\rho} 
\left ( \nu \nabla^2 {\bf U}_g + D \frac{{\bf V}}{\rho} \nabla^2 \rho_p \right )  +  \\ \frac{\rho_g}{\rho} 
\left [ \nu \nabla^2 {\bf u}_g + D \frac{{\bf V}}{\rho} \nabla^2\rho_p^\prime + 
D \left ( {\bf v} - \frac{\rho_p^{\prime}}{\rho} {\bf V} \right )
\frac{\nabla^2\rho_p}{\rho}  \right ],
\end{aligned}
\end{equation}

\begin{equation}
\label{G_nu}
{\bf G}_\nu = -\nu \frac{\rho_g}{\rho} \frac{\rho_p^\prime}{\rho} \nabla^2 {\bf U}_g + \nu \frac{\rho_g}{\rho} \nabla^2 {\bf u}_g
\end{equation}
and 
\begin{equation}
\label{pert_u_g}
{\bf u}_g = {\bf u} - \frac{\rho_p}{\rho} {\bf v} - 
\frac{\rho_g}{\rho} \frac{\rho_p^\prime}{\rho} {\bf V}.
\end{equation}

Eqs. (\ref{pert_U}-\ref{pert_rho_tot}) contain the Eulerian perturbations of ${\bm U}_g$, ${\bm U}$, ${\bm V}$, $\rho_p$ and $p$,
which are, 
respectively, ${\bm u}_g$, ${\bm u}$, ${\bm v}$, $\rho_p^\prime$ and $p^\prime$.
Eqs. (\ref{pert_U}-\ref{pert_rho_tot}) recover the corresponding inviscid equations (A1-A4) given in \citetalias{zhuravlev-2019}.

The background specified by eqs. (\ref{bg_U}-\ref{bg_sigma}) omitting the term $\sim t_s^2$ in eq. (\ref{bg_V}) 
yields the following equations for ${\bm u}$, ${\bm v}$ and $\delta$
\begin{equation}
\label{u}
\begin{aligned}
{\cal D} {\bf u} - 2\Omega_0 u_y {\bf e}_x + 
(2-q)\Omega_0 u_x {\bf e}_y =  -\nabla W +\frac{f}{1+f} \delta {\bf g} + \\
\frac{1}{1+f} \left [ \nu \left ( \nabla^2 {\bf u} - \frac{f}{1+f} \nabla^2 {\bf v} \right ) - \frac{f \tau {\bf g}}{(1+f)} 
\nabla^2 \delta\, \left ( \frac{\nu}{1+f} - D \right ) \right ],
\end{aligned}
\end{equation}
\begin{equation}
\label{v}
\begin{aligned}
\left [ 1 - \frac{f}{(1+f)^2} \tau \nu \nabla^2 \right ] \frac{{\bf v}}{\tau} = \nabla W - \frac{1}{1+f} \nu \nabla^2 {\bf u}\, - \\ 
\frac{f}{1+f} \left [ 1 - \frac{1}{(1+f)^2} \tau\nu\nabla^2 \right ] \delta\, {\bf g}, 
\end{aligned}
\end{equation}
\begin{equation}
\label{delta}
{\cal D} {\delta} = -\tau \nabla^2 W - \frac{1-f}{1+f} \tau ({\bf g} \nabla) \delta + D \nabla^2 \delta,
\end{equation}

\begin{equation}
\label{div_u}
\nabla \cdot {\bf u} = \frac{\tau f}{1+f} \nabla^2 W + \tau f \frac{1-f}{(1+f)^2} ({\bf g} \nabla) \delta,
\end{equation}
where perturbations of pressure and dust density are replaced by 
$W \equiv p^\prime/\rho$ and $\delta \equiv \rho_p^\prime/\rho_p$, respectively, while it is taken into account 
that $\nabla \cdot {\bm G}_\nu = 0$. 
Eqs. (\ref{u}-\ref{div_u}) recover eqs. (A5-A7) of \citetalias{zhuravlev-2019} in the inviscid limit.

In their general form, eqs. (\ref{u}-\ref{div_u}) are hardly tractable analytically. 
However, it might be instructive 
to consider the limiting case over the main parameter of the problem being the stopping time in units of the characteristic time 
of viscous damping, $\tau_{\nu}$. 
As $\tau_{\nu}\ll 1$, the viscous terms in square brackets of eq. (\ref{v}) can be omitted, which makes it possible
to see that, within the same approximation, the term containing ${\bf v}$ in eq. (\ref{u}) gives a small contribution to RHS of
eq. (\ref{u}) in comparison with the inviscid terms therein. Additionally, the viscous term $\sim \nabla^2 \delta$ in the
square brackets in RHS of eq. (\ref{u}) can be omitted by the same reasoning. Thus, in the limit $\tau_\nu\ll 1$ 
eqs. (\ref{u}-\ref{div_u}) become similar to the corresponding inviscid equations 
with the only addition of
the term containing $\nabla^2 {\bm u}$ in RHS of eq. (\ref{u}) and the term containing $\nabla^2 \delta$ in RHS of eq. (\ref{delta}).

\subsection{The limit of the small dust fraction}

In the limit of the small dust fraction, $f\ll 1$, eqs. (\ref{u}-\ref{div_u}) 
reduce to the following set of equations for ${\bf u}$, $W$ and $\delta$.

\begin{equation}
\label{u_x}
{\cal D} u_x - 2\Omega_0 u_y = - \partial_x W - f g_x \delta + \nu \nabla^2 u_x,
\end{equation}
\begin{equation}
\label{u_y}
{\cal D} u_y + \frac{\kappa^2}{2\Omega_0} u_x = - \partial_y W + \nu \nabla^2 u_y,
\end{equation}
\begin{equation}
\label{u_z}
{\cal D} u_z = - \partial_z W - f g_z \delta + \nu \nabla^2 u_z,
\end{equation}
\begin{equation}
\label{delta_2}
{\cal D} \delta = - \tau \nabla^2 W + \tau (g_x \partial_x + g_z \partial_z) \delta + D \nabla^2 \delta,
\end{equation}
\begin{equation}
\label{div_u_2}
\partial_x u_x + \partial_y u_y + \partial_z u_z = 0.
\end{equation}
The system (\ref{u_x}-\ref{div_u_2}) reproduces eqs. (A8-A12) of \citetalias{zhuravlev-2019} in the inviscid limit 
with an additional condition of axial symmetry of perturbations.


Similar to the case of the inviscid dynamics, 
in the basic model considered here,
the small gas-dust perturbations of dust streaming through the turbulent gas environment 
are described by eqs. (\ref{u_x}-\ref{u_z}, \ref{div_u_2}) for perturbation 
of the centre-of-mass velocity, ${\bf u}$, which recover equations for dynamics 
of vortical single-fluid perturbations in a viscous rotating plane shear flow, see e.g. \citet{landau-lifshitz-1987} paragraph 14,
with an additional terms in RHS of eqs. (\ref{u_x}) and (\ref{u_z}) accounting for the dust back reaction on gas. 
Perturbation of the dust density is governed by separate eq. (\ref{delta_2}), which accounts for the dust turbulent diffusion.

\end{document}